\documentclass[aps,prd,onecolumn,nofootinbib]{revtex4-2}
\pdfpagewidth=\paperwidth
\pdfpageheight=\paperheight
\usepackage{graphicx}%Include figure files
\usepackage{subfigure}
\usepackage{amsmath}
\usepackage{amsfonts}
\usepackage{amssymb}
\usepackage{subfigure}
\usepackage{color}%
\usepackage{dcolumn}% Align table columns on decimal pointl

\usepackage{MnSymbol,wasysym}
\usepackage{braket}
\usepackage{eurosym}
\usepackage{calrsfs}
\usepackage{multirow}
\usepackage[colorlinks=true,linkcolor=blue,citecolor=blue]{hyperref}
\usepackage[title]{appendix}
\begin{document}
\baselineskip=0.5 cm

\title{Mapping Quasi-Periodic Oscillations to Lyapunov Exponent across Black Hole Thermodynamic Phase Transitions}

\author{R. H. Ali}
\email{hasnainali408@yzu.edu.cn,hasnainali408@gmail.com}
\affiliation{Center for Gravitation and Cosmology, College of Physical Science and Technology, Yangzhou University, Yangzhou, 225009, China}

\author{Zi-Yu Tang}
\email{tangziyu@ibs.re.kr (Corresponding author)}
\affiliation{Cosmology, Gravity and Astroparticle Physics Group, Center for Theoretical Physics of the Universe, Institute for Basic Science, Daejeon 34126, Korea}

\author{Xiao-Mei Kuang}
\email{xmeikuang@yzu.edu.cn (Corresponding author)}
\affiliation{Center for Gravitation and Cosmology, College of Physical Science and Technology, Yangzhou University, Yangzhou, 225009, China}

\begin{abstract}
\baselineskip=0.5 cm
We investigate the thermodynamic phase structure of a nonminimally coupled magnetic AdS black hole through the dynamics of timelike particles. The free energy analysis reveals a Van der Waals-like phase transition characterized by small, intermediate, and large black hole phases. We show that both the Lyapunov exponent of \textit{unstable} circular orbits and the quasi-periodic oscillation (QPO) frequencies associated with \textit{stable} circular orbits exhibit clear signatures of underlying thermodynamic phase structure, including first order and critical phase transitions. More importantly, we establish a QPO–Lyapunov exponent mapping and demonstrate that the resulting relation inherits the same thermodynamic branch structure. Although the Lyapunov exponent and QPO frequencies originate from unstable and stable circular orbits, respectively, their correspondence emerges from the common black hole spacetime geometry and remains valid even in the absence of phase transitions. Our results reveal an unexplored connection among orbital instability, QPO phenomenology, and black hole thermodynamics, suggesting a potential observational route for probing chaotic orbital dynamics and thermodynamic phases through QPO measurements.
\end{abstract}

\maketitle
\newpage

\textit{}\tableofcontents

\section{Introduction}
Black hole provides a unique laboratory for exploring gravity in the strong field regime and testing General Relativity (GR) under extreme conditions. The advent of gravitational wave astronomy, marked by the first detection of a binary black hole merger by the LIGO-Virgo Collaborations~\cite{LIGOScientific:2016aoc}, together with the horizon scale images of the supermassive black holes M87* and SgrA* reconstructed by the Event Horizon Telescope~\cite{EventHorizonTelescope:2019dse,EventHorizonTelescope:2022wkp}, has opened a new era of black hole observations. These breakthroughs have greatly stimulated efforts to identify observable signatures of black hole spacetimes and their underlying physical properties.

Black hole thermodynamics provide one of the deepest connections between gravitation, quantum theory, and statistical physics. Following the discovery of the Bekenstein-Hawking entropy~\cite{Bekenstein:1973ur,Hawking:1975vcx}, the formulation of the four laws of black hole mechanics~\cite{Bardeen:1973gs}, and Wald's geometric interpretation of black hole entropy~\cite{Wald:1979zz,Wald:1999vt}, black holes have been recognized as genuine thermodynamic systems. A major milestone was the Hawking-Page phase transition in asymptotically anti-de Sitter (AdS) spacetime~\cite{Hawking:1982dh}, demonstrating that black holes can undergo thermodynamic phase transitions. Later, treating the cosmological constant as thermodynamic pressure extended black hole thermodynamics into the framework of black hole chemistry~\cite{Dolan:2010ha,dolan2012pdv,Kubiznak:2016qmn}, within which rich phase structures analogous to ordinary thermodynamic systems, most notably the Van der Waals-like phase transition, have been extensively explored~\cite{Kubiznak:2012wp}. Extensive studies within this framework have revealed rich thermodynamic phenomena in a broad class of black hole solutions, including a number of models investigated by some of us~\cite{Fang:2017nse,Hu:2018prt,Kuang:2018goo,Cisterna:2018jqg,Xing:2021gpn,Ladino:2024ned}. These developments naturally raise the question of how such thermodynamic phase structures may manifest themselves through observable or dynamical quantities.

Considerable effort has therefore been devoted to identifying dynamical and observational probes of black hole thermodynamics. Besides conventional thermodynamic quantities and the Ruppeiner geometry~\cite{Ruppeiner:2012uc,Wei:2019yvs,Miao:2017cyt,Wei:2019uqg}, various physical observables have been shown to encode information about thermodynamic phase transitions, including quasinormal modes~\cite{Liu:2014gvf,Mahapatra:2016dae}, circular geodesics~\cite{Wei:2017mwc,Zhang:2019tzi}, and black hole shadows~\cite{Zhang:2019glo}. In particular, more recently, the Lyapunov exponent associated with unstable circular orbits has attracted more attention. It has been shown that the Lyapunov exponent can effectively characterize black hole phase transitions and even exhibits the universal critical exponent $1/2$ near the critical point~\cite{Guo:2021enm,Chen:2022qrw,Weng:2025fib,Yang:2023hci,Lyu:2023sih,Kumara:2024obd,Du:2024uhd,Shukla:2024tkw,Gogoi:2024akv,Chen:2025xqc,R:2025gok,Bezboruah:2025udi,Ali:2025ooh}.

The Lyapunov exponent characterizes the instability timescale of unstable circular orbits and therefore provides a direct measure of orbital dynamics in black hole spacetimes. Owing to its geometric origin, it has found broad applications in black hole physics, ranging from orbital stability analyses to the study of black hole perturbations. In particular, it was shown that, in the eikonal limit, the quasinormal modes of black holes are closely related to the properties of unstable null circular geodesics, with the Lyapunov exponent determining the corresponding instability timescale~\cite{Cardoso:2008bp}. More recently, the Lyapunov exponent of timelike circular orbits has emerged as a promising dynamical probe of black hole thermodynamic phase transitions, establishing an intriguing connection between orbital instability and black hole thermodynamics~\cite{Guo:2021enm,Chen:2022qrw,Weng:2025fib,Yang:2023hci,Lyu:2023sih,Kumara:2024obd,Du:2024uhd,Shukla:2024tkw,Gogoi:2024akv,Chen:2025xqc,R:2025gok,Bezboruah:2025udi,Ali:2025ooh}.

Another important observational probe of black hole spacetimes is provided by quasi-periodic oscillations (QPOs), which appear as narrow peaks in the X-ray power spectra of accreting compact objects~\cite{Lewinbook,Motta:2016vwf,eXTP:2018kpm}. Among the various theoretical models proposed to explain their origin, the relativistic precession (RP) model relates the observed QPO frequencies to the orbital and epicyclic frequencies of test particle motion in strong gravitational fields~\cite{Kluzniak1990,Stella:1997tc,Stella:1999sj}. Since these characteristic frequencies are determined directly by the underlying spacetime geometry, QPOs have been widely employed to constrain black hole parameters and to test gravity in the strong field regime~\cite{Zhang:2009gn,Bambi:2012pa,Maselli:2014fca,Jusufi:2020odz,Ghasemi-Nodehi:2020oiz,Chen:2021jgj,Allahyari:2021bsq,Deligianni:2021ecz,Jiang:2021ajk,Banerjee:2022chn,Liu:2023vfh,Riaz:2023yde,Rayimbaev:2023bjs,Abdulkhamidov:2024lvp,Jumaniyozov:2024eah,Guo:2025zca,Wu:2025ccc,Wu:2025xtn}. Motivated by the success of dynamical probes of black hole thermodynamics, recent studies have also begun to explore whether QPO frequencies carry signatures of thermodynamic phase transitions~\cite{Hazarika:2025zgv}.

However, despite the rapid development of both directions, the relationship between the Lyapunov exponent and QPO frequencies remains largely unexplored. Although both quantities originate from geodesic motion in black hole spacetimes, they are associated with fundamentally different classes of circular orbits: the Lyapunov exponent characterizes the instability of unstable circular orbits, whereas QPO frequencies are generated by small oscillations around stable circular orbits. Whether these seemingly distinct dynamical quantities share common information about the underlying black hole spacetime and its thermodynamic properties therefore remains an open question.

In this work, we investigate these questions in the background of a nonminimally coupled magnetic AdS black hole, described by an exact solution of the Einstein-Yang-Mills theory with an $SU(2)$ Wu-Yang magnetic gauge field~\cite{Balakin:2006gv}. Nonminimal theories, in which the gravitational field couples to matter fields through spacetime curvature, have attracted considerable attention as extensions of Einstein gravity~\cite{scherrer1941theorie,jordan1945projektiven,Brans:1961sx,Hehl:1999bt,Mueller-Hoissen:1987nvb,Balakin:2007mp,Balakin:2009rg}. Within this framework, a variety of exact solutions have been constructed using the $SU(2)$ Wu-Yang ansatz, including electric and magnetic black holes, stars, and wormholes~\cite{Horndeski:1978ca,Mueller-Hoissen:1988cpx,Balakin:2007xq,Balakin:2007am,Balakin:2015oea}, whose physical properties have been extensively investigated~\cite{vanderBij:2000cu,Ayon-Beato:2000mjt,Horvat:2004qs,Bronnikov:2006fu,Uchikata:2012zs,Aftergood:2014wla,Ma:2015gpa}.

We show that both the Lyapunov exponent and the QPO frequencies faithfully encode the underlying thermodynamic phase structure of the black hole. More importantly, we establish a direct mapping between these two dynamical quantities and demonstrate that this correspondence inherits the same thermodynamic branch structure. Although the Lyapunov exponent and the QPO frequencies originate from unstable and stable circular orbits, respectively, their correspondence is governed by the common spacetime geometry and remains valid even in the absence of thermodynamic phase transitions. Our results therefore reveal a previously unexplored connection among orbital instability, QPO phenomenology, and black hole thermodynamics, providing a new dynamical perspective on black hole phase transitions.

The paper is organized as follows. In Sec.~\ref{sec:Background}, we review the thermodynamic properties and phase structure of the nonminimally coupled magnetic AdS black hole. In Sec.~\ref{sec:Time-like LE}, we investigate the geodesic motion of massive particles and analyze how the Lyapunov exponent encodes the underlying thermodynamic phase structure. In Sec.~\ref{sec:QPO phase transition}, we study the QPO frequencies within the RP model and examine their behavior across the thermodynamic phase transition, before establishing their correspondence with the Lyapunov exponent in Sec.~\ref{Mapping QPO–Lyapunov Exponent Across Thermodynamic Phases}. Finally, we summarize our main results and discuss their implications in Sec.~\ref{sec:conclusion}.

\section{Thermodynamic Phase Structure of a Nonminimally Coupled Magnetic AdS Black Hole} \label{sec:Background}
The nonminimally coupled Einstein-Yang-Mills theory in an AdS spacetime is described by the action \cite{Balakin:2015gpq}
\begin{equation}
\mathbf{S}_{\text{NMEYM}}=\frac{1}{2}\int d^{4}x\sqrt{-g}\Big(\frac{R-2\Lambda}{8\pi G}-\frac{1}{2}F^{(\alpha)}_{jk}F^{jk(\alpha)}+ \frac{1}{2}\mathfrak{R}^{jkpn}F^{(\alpha)}_{jk}F^{(\alpha)}_{pn}\Big).\label{Eq:action}
\end{equation}
Here, $g=\text{det}[g_{jk}]$ and $R$ represent the determinant of the metric tensor and Ricci scalar, respectively. $\Lambda=-6/l^2$ is the cosmological constant with the AdS radius $l$. We use geometrized units $c=G=1$ unless otherwise specified. Noted that the Latin in the action indices range from $0$ to $3$, while the Greek indices  ranges from $1$ to $3$. The $SU(2)$ Yang-Mills field strength $F^{(\alpha)}_{jk}$ is related with a triplet of vector potentials $A^{(\alpha)}_j$ through the formulas
\begin{equation}
F^{(\alpha)}_{jk} = \nabla_j A^{(\alpha)}_k - \nabla_k A^{(\alpha)}_j + f^{(\alpha)}_{(\rho)(\sigma)} A^{(\rho)}_j A^{(\sigma)}_k, \label{eq:potential}
\end{equation}
where $\nabla_j$ is the covariant derivative and $f^{(\alpha)}_{(\rho)(\sigma)}$ represents the real structure constants of the $3$-parameter Yang-Mills gauge group $SU(2)$.
The nonminimal susceptibility tensor $\mathfrak{R}^{jkpn}$ is defined as
\begin{equation}
\mathfrak{R}^{jkpn}=\frac{q_{1}}{2}R\Big(g^{jp}g^{kn}-g^{jn}g^{kp}\Big)+\frac{q_{2}}{2}\Big(R^{jp}g^{kn}-R^{jn}g^{kp}+R^{kn}g^{jp}-R^{kp}g^{jn}\Big)+q_{3}{R}^{jkpn},
\label{Eq:tensor}
\end{equation}
where $R^{jk}$, $R^{jkpn}$ are the Ricci tensor and Riemannian tensor, respectively, while $q_{1}, q_{2}, q_{3}$ represent the phenomenological parameters describing the nonminimal coupling between the Yang-Mills field and the gravitational field.
It was addressed in \cite{Balakin:2015gpq} that by setting $q_{1}=-\xi, q_{2}=4\xi$ and $q_{3}=-6\xi$, the equations of motion derived from above action admits an analytical static spherically symmetric black hole solution
\begin{equation}
    ds^{2}= -N(r)dt^{2}+ \frac{1}{N(r)}dr^{2}+ r^{2}(d\theta^{2}+\sin^{2}\theta d\phi^{2}),
\end{equation}
with the metric function
\begin{equation}
    N(r)=1+\Big(\frac{r^4}{r^4+2 \xi  {Q_m}^2}\Big )\left(-\frac{2 M}{r}+\frac{{Q_m}^2}{r^2}+\frac{r^2}{l^2}\right)~. \label{Eq:metric}
\end{equation}
Here, $\xi$ plays the role of nonminimal coupling parameter. It is worth noting that for $\xi>0$ the solution describes a regular black hole without curvature singularities. Moreover, the magnetic charge ${Q_m}$  is associated with the Wu–Yang $SU(2)$ gauge configuration and subject to topological quantization \cite{wu1969some, shnir2005magnetic,Balakin:2006gv}
\begin{equation}
    Q_m^2 \equiv 4\pi \nu^2~,
\end{equation}
where $\nu$ is a nonvanishing integer. However, in the following thermodynamic analysis, we will treat  $Q_m$ as an effective continuous parameter at the classical level.

As discussed in \cite{Balakin:2015gpq}, for asymptotically AdS spacetimes with $\Lambda>0$, the metric function (\ref{Eq:metric}) admits at most two positive roots, corresponding to the Cauchy horizon ($r_-$) and the event horizon ($r_+$). When the two horizons coincide, the black hole becomes extremal; otherwise, no black hole solution exists.

\begin{figure}[ht!]
\centering
\includegraphics[width=7.5cm]{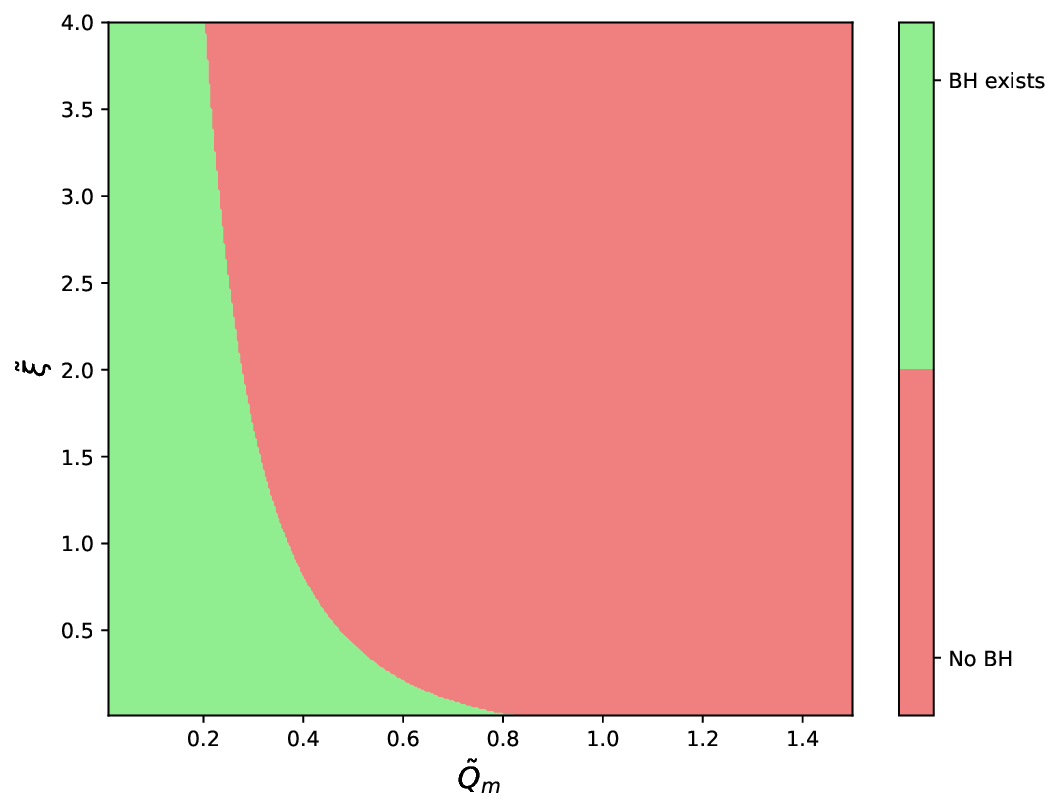}\hspace{1cm}
\caption{Typical phase diagram in the $\tilde{Q}_m-\tilde{\xi}$ parameter space, showing the regions admitting black hole solutions and horizonless configurations, thereby delineating the physically allowed domain of nonminimal coupled magnetic AdS black holes. We fix $\tilde{M}=1$ and show only $\tilde{Q}_m\ge0$, as the diagram is symmetric under $\tilde{Q}_m \to -\tilde{Q}_m$.}\label{fig:space-parameter}
\end{figure}

To characterize the parameter space, we introduce the following dimensionless quantities 
\begin{equation}
    \tilde{M}\equiv M/l~, \quad \tilde{Q}_m\equiv Q_m/l~, \quad \tilde{r}_h \equiv r_h/l~, \quad \tilde{\xi}=\xi/l^2~,
\end{equation}
where $r_h$ denotes the horizon radius, corresponding to the positive roots of the metric function, namely the Cauchy horizon $r_-$ or the event horizon $r_+$. The horizon condition $N(r_h)=0$ can then be rewritten as
\begin{equation}
    \tilde{r}_h^6+\tilde{r}_h^4-2 \tilde{M} \tilde{r}_h^3+\tilde{Q}_m^2 \left(2 \tilde{\xi}+\tilde{r}_h^2\right)=0~. \label{M_relation}
\end{equation}
This equation explicitly shows that the dimensionless horizon radius $\tilde{r}_h$ is independent of the AdS radius $l$. Equivalently, the ratio between the horizon radius and the AdS radius is fully determined by the dimensionless parameters $\tilde{M}$, $\tilde{Q}_m$ and $\tilde{\xi}$. Moreover, we introduce two auxiliary functions, 
\begin{eqnarray}
    H_1(\tilde{r})&=&\tilde{r}^6+\tilde{r}^4-2 \tilde{M}\tilde{r}^3+\tilde{Q}_m^2 \left(2 \tilde{\xi} +\tilde{r}^2\right)~, \\
    H_2(\tilde{r})&=&\frac{1}{2\tilde{r}}H_1'(\tilde{r})=\tilde{r} \left(3 \tilde{r}^3-3 \tilde{M}+2 \tilde{r}\right)+\tilde{Q}_m^2~.   
\end{eqnarray}
The existence of asymptotically AdS black hole solutions requires the conditions $H_2(\tilde{r}_0)<0$ and $H_1(\tilde{r}_2)\le0$. Here, $\tilde{r}_0$ denotes the unique positive root of $H_2'(\tilde{r})=0$, while $\tilde{r}_2$ is the larger positive root of $H_2(\tilde{r})=0$. Since the explicit expressions of these conditions are complicated, we fix $\tilde{M}=1$ to illustrate a typical phase diagram, shown in Fig.~\ref{fig:space-parameter}, in the $\tilde{Q}_m-\tilde{\xi}$ parameter space, which delineates regions admitting black hole solutions and horizonless spacetimes.

\begin{figure}[ht!]
\centering
\includegraphics[width=7cm]{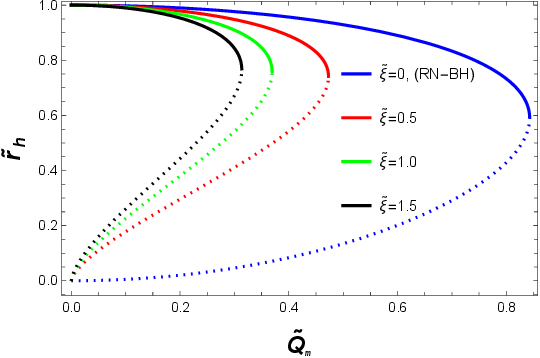}\hspace{1cm}
\caption{Typical plots showing the dependence of the horizon radius on the magnetic charge $\tilde{Q}_m$ for various values of the dimensionless coupling parameter $\tilde{\xi}$. The solid curves represent the event horizon radius ($\tilde{r}_+$), while the dotted curves correspond to the Cauchy horizon ($\tilde{r}_-$). Their intersection indicates the extremal black hole case. We set $\tilde{M}=1$.}\label{Fig:horizon1}
\end{figure}

Our subsequent analysis focuses on the parameter regime that admits nonminimally coupled magnetic AdS black hole solutions. The dependence of the dimensionless horizon radius $\tilde{r}_h$, namely $\tilde{r}_-$ and $\tilde{r}_+$, on the magnetic charge $\tilde{Q}_m$ for selected values of the nonminimal coupling parameter is illustrated in Fig.~\ref{Fig:horizon1}. We observe that larger values of $\tilde{Q}_m$ and $\tilde{\xi}$ correspond to a smaller event horizon radius and a larger Cauchy horizon radius.

%%%%%%%%%%%
%%%%%%%%%%%
%%%%%%%%%%
%%%%%%%%%%%

%%%%%%%%%%

In the standard geometric approach, the Hawking temperature is determined by the periodicity of the Euclidean time coordinate at the event horizon
\begin{equation}
T=\frac{1}{4\pi}N'(r_+)=-\frac{6  \xi  l^2 {Q_m}^2+l^2 {Q_m}^2 r_+^2-l^2 r_+^4-3 r_+^6}{8 \pi   \xi  l^2 {Q_m}^2 r_++4 \pi  l^2 r_+^5}~,\label{Eq:temperature}
\end{equation}
where we have used the relation
\begin{equation}
M=\frac{\xi  {Q_m}^2}{r_+^3}+\frac{{Q_m}^2}{2 r_+}+\frac{r_+^3}{2l^2}+\frac{r_+}{2} \label{Eq:mass}
\end{equation}
which follows from the horizon condition $N(r_+)=0$.

Also, the thermal entropy of the black hole can be obtained following the Wald entropy formula \cite{Wald:1993nt,Iyer:1994ys, Bodendorfer:2013wga}
\begin{eqnarray}
    S&=&-2\pi \int_{\Sigma}\frac{\partial \mathfrak{L}}{\partial R_{\mu\nu\rho\sigma}}\epsilon_{\mu\nu}\epsilon_{\rho\sigma}\sqrt{h}d^2x~\notag\\
    &=&-2\pi \int_{\Sigma}\left[-2\left(\frac{1}{16\pi G}+\frac{q_1}{4}F_{mn}F^{mn}\right)+\frac{q_2}{4}\left(F^\mu{}_n F^{\rho n}+F^\rho{}_n F^{\mu n}\right)+\frac{q_3}{2}F^{\mu\nu}\epsilon_{\mu\nu}F^{\rho\sigma}\epsilon_{\rho\sigma}\right]\sqrt{h}d^2x~\notag\\
    &=&4\pi \left(\frac{1}{16\pi}+\frac{q_1}{2}\frac{\nu^2}{r_+^4}\right)4\pi r_+^2~\notag \\
    &=&\pi r_+^2-2\pi \xi \frac{Q_m^2}{r_+^2}~,\label{Eq:entropy}
\end{eqnarray}
where $\mathfrak{L}$ is the Lagrangian density given in Eq.~\eqref{Eq:action}. It is noteworthy that the regularity condition $\xi>0$ leads to a reduction of the entropy compared with the standard area law. This correction originates from the nonminimal  coupling between the matters and curvature, and reflects the nontrivial contribution of the magnetic field to the gravitational degrees of freedom. In contrast, for $\xi<0$, corresponding to a singular black hole spacetime, the entropy increases with the magnetic charge. Notably, for sufficiently small black holes with the radius of event horizon $r_+<\left(2\xi Q_m^2\right)^{1/4}$, the entropy can become negative. The requirement of non-negative entropy implies the existence of a minimum horizon radius $r_+^{\rm min}=\left(2\xi Q_m^2\right)^{1/4}$, below which physically admissible black hole solutions are excluded. Nevertheless, for a regular black hole with $\xi>0$, the entropy increases monotonically with the horizon radius $r_+$. In contrast, for a singular black hole with $\xi<0$, the entropy exhibits a minimum entropy $S_{\rm min}=2\pi \sqrt{-2\xi Q_m^2}$ attained at the horizon radius $r_+=\left(-2\xi Q_m^2\right)^{1/4}$. This radius coincides with the location of the curvature singularity. Therefore, physically admissible black hole solutions must satisfy $r_+>\left(-2\xi Q_m^2\right)^{1/4}$, and the entropy then increases monotonically with the horizon radius.

We next investigate the thermodynamic properties in the canonical ensemble with fixed magnetic charge. Since the magnetic charge in the present theory is a conserved and quantized topological quantity and cannot be exchanged with the environment, the appropriate thermodynamic potential is the Helmholtz free energy 
\begin{equation}
F\equiv M-TS=\frac{-4 l^2 \xi ^2 Q_m^4+r_+^6 \left(3 l^2+10 \xi \right) Q_m^2+16 l^2 \xi  r_+^4 Q_m^2+2 l^2 \xi  r_+^2 Q_m^4+l^2 r_+^8-r_+^{10}}{4 l^2 r_+^3 \left(2 \xi  Q_m^2+r_+^4\right)}.\label{Eq:free energy}
\end{equation}

These thermodynamic quantities can be further rescaled into dimensionless form as 
\begin{eqnarray}
    &&\tilde{T}\equiv T l=-\frac{6 \tilde{\xi} \tilde{Q}_m^2+\tilde{Q}_m^2 \tilde{r}_+^2-3 \tilde{r}_+^6-\tilde{r}_+^4}{8 \pi  \tilde{\xi} \tilde{Q}_m^2 \tilde{r}_++4 \pi  \tilde{r}_+^5}~,\\
    &&\tilde{S}\equiv S/l^2=\frac{\pi  \left(\tilde{r}_+^4-2 \tilde{\xi} \tilde{Q}_m^2\right)}{\tilde{r}_+^2}~,\\
    &&\tilde{F}\equiv F/l=\frac{2 \tilde{\xi} \tilde{Q}_m^4 \left(\tilde{r}_+^2-2 \tilde{\xi}\right)+\tilde{Q}_m^2 \left((10 \tilde{\xi}+3) \tilde{r}_+^6+16 \tilde{\xi} \tilde{r}_+^4\right)-\tilde{r}_+^{10}+\tilde{r}_+^8}{4 \tilde{r}_+^3 \left(2 \tilde{\xi} \tilde{Q}_m^2+\tilde{r}_+^4\right)}~,
\end{eqnarray}
where the functional relations among the rescaled thermodynamic quantities are independent of the AdS length scale $l$.

A theoretical framework contextualizes the study of phase transitions in AdS black hole models by determining the thermodynamic critical quantities and probing the influence of black hole parameters and nature of the phase transitions. In this regard, a key quantity is the Hawking temperature, as it governs the black hole thermodynamic properties and dictates the nature of its phase transitions. The utilization of the horizon-temperature plane is significant for studying the critical behavior, and one can establish the existence of a characteristic critical point. Therefore, criticality is identified by an inflection point in the curve, defined by the simultaneous satisfaction of the following conditions
\begin{equation}
\left(\frac{\partial \tilde{T}}{\partial \tilde{r}_{+}}\right)_{\mathrm{cri}} = 0,
\quad \mathrm{and} \quad \left(\frac{\partial^{2}\tilde{T}}{\partial \tilde{r}_{+}^{2}}\right)_{\mathrm{cri}} = 0.\label{Eq:cri}
\end{equation}
Consequently, the present investigation focuses on the phase structure of the magnetic AdS black hole described by Eq.~\eqref{Eq:metric}, and a comprehensive analysis reveals a rich landscape of thermodynamic phase transitions.

The critical values are obtained by simultaneously solving the inflection point conditions given in Eq.~\eqref{Eq:cri}. For a given value of the dimensionless magnetic charge $\tilde{Q}_m$, the critical event horizon radius $\tilde{r}_+^c$ and coupling parameter $\tilde{\xi}_c$ are determined from these conditions, and the corresponding critical Hawking temperature $\tilde{T}_c$ is subsequently obtained. However, due to the complicated structure of the equations, analytical solutions are generally inaccessible, which necessitates the use of robust numerical techniques to accurately characterize the critical landscape.

\begin{figure}[ht!]
\centering
{\includegraphics[width=6cm]{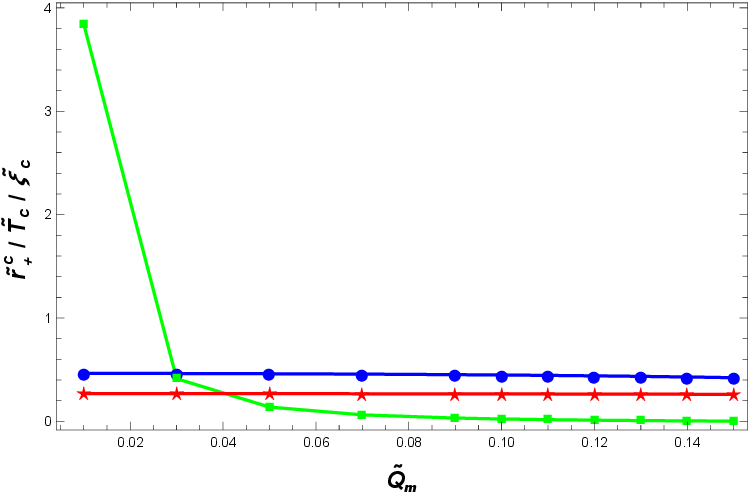}}\hspace{1cm}
{\includegraphics[width=6cm]{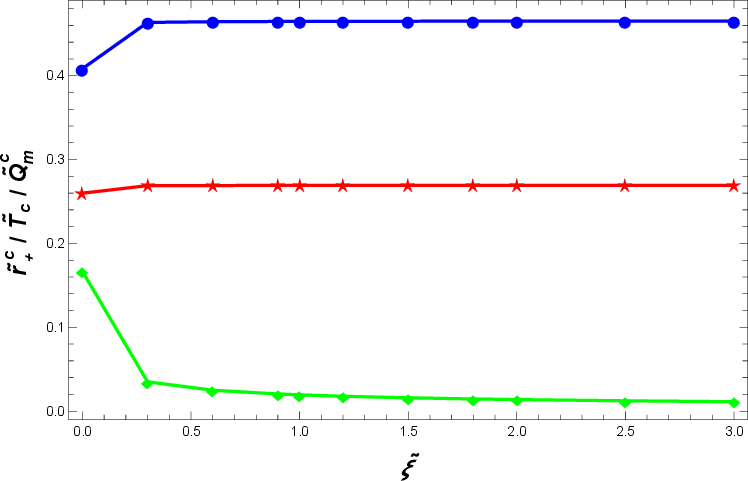}}
\caption{The critical dimensionless event horizon radius $\tilde{r}_+^c$ (blue), Hawking temperature $\tilde{T}_c$ (red), and the coupling parameter $\tilde{\xi}_c$ (green) or magnetic charge $\tilde{Q}_m$ (green) are plotted as functions of $\tilde{Q}_m$ and $\tilde{\xi}$ in the left and right panels, respectively.}\label{fig:critical point}
\end{figure}

The resulting critical points for selected parameter values are summarized in Tables~\ref{table:Fix-Q-cri} and~\ref{table:Fix-lamda-cri}, while their overall trends are visualized in Fig.~\ref{fig:critical point}. Specifically, the left (right) panel shows the dependence of the critical points on $\tilde{Q}_m$ ($\tilde{\xi}$), providing a clear graphical illustration of how the critical behavior evolves with the black hole parameters. It is shown that as $\tilde{Q}_m$ ($\tilde{\xi}$) increases, the critical coupling parameter $\tilde{\xi}_c$ (critical magnetic charge $\tilde{Q}_m^c$) drops rapidly, whereas the critical dimensionless event horizon radius $\tilde{r}_+^c$ and the Hawking temperature $\tilde{T}_c$ decrease (increase) more gently.

\begin{table}[ht!]
\centering % used for centering table
\begin{tabular}{c c c c c c c c c c } % centered columns (10 columns for alignment)
 \hline %inserts double horizontal lines
${\tilde{Q}_m}$& & ${\tilde{r}_+^c}$ &~~~~~~~~~~~~~ &${\tilde{\xi}}_{c}$ &  ~~~~~~~~~~~~~~~~~~~&~${\tilde{T}_{c}}$ &~~~~~~ & \\ [0.5ex]
        \hline
$0.01$ ~~~~&~~ &$0.46511$& ~~~~&~~ $3.8420$& ~~~~~&~~ $0.26920$&~~~~~~~~~~\\
$0.03$ ~~~~&~~ &$0.46393$& ~~~~&~~ $0.41347$& ~~~~~&~~ $0.26897$&~~~~~~~~~~\\
$0.05$ ~~~~&~~ &$0.46152$& ~~~~&~~ $0.13925$& ~~~~~&~~ $0.26852$&~~~~~~~~~~\\
$0.07$ ~~~~&~~ &$0.45778$& ~~~~&~~ $0.063773$& ~~~~&~~ $0.26782$&~~~~~~~~~~\\
$0.09$ ~~~~&~~ &$0.45250$& ~~~~&~~ $0.032802$& ~~~~&~~ $0.26686$&~~~~~~~~~~\\
$0.10$ ~~~~&~~ &$0.44921$& ~~~~&~~ $0.023832$& ~~~~&~~ $0.26627$&~~~~~~~~~~\\
$0.11$ ~~~~&~~ &$0.44541$& ~~~~&~~ $0.017226$& ~~~~&~~ $0.26561$&~~~~~~~~~~\\
$0.12$ ~~~~&~~ &$0.44103$& ~~~~&~~ $0.012230$& ~~~~&~~ $0.26485$&~~~~~~~~~~\\
$0.13$ ~~~~&~~ &$0.43596$& ~~~~&~~ $0.0083800$& ~~~~&~~ $0.26401$&~~~~~~~~~~\\
$0.14$ ~~~~&~~ &$0.43006$& ~~~~&~~ $0.0053700$& ~~~~&~~ $0.26306$&~~~~~~~~~~\\
$0.15$ ~~~~&~~ &$0.42313$& ~~~~&~~ $0.0029700$& ~~~~&~~ $0.26200$&~~~~~~~~~~\\
\hline
\end{tabular}
\caption{The critical values of the dimensionless event horizon radius $\tilde{r}_+^c$, coupling parameter $\tilde{\xi}_c$, and Hawking temperature $\tilde{T}_c$ for selected values of the dimensionless magnetic charge $\tilde{Q}_m$ of the black hole.
} % title of Table
\label{table:Fix-Q-cri}
\end{table}

%%%%%%%%%%
%%%%%%%%%%
\begin{table}[ht!]
\centering % used for centering table
\begin{tabular}{c c c c c c c c c c } % centered columns (7 columns)
      \hline %inserts double horizontal lines
${\tilde{\xi}}$& & ${\tilde{r}_+^c}$ &~~~~~~~~~~~~~ &$\tilde{Q}_m^c$ &  ~~~~~~~~~~~~~~~~~~~&~${\tilde{T}_{c}}$ &~~~~~~ & \\ [0.5ex]
                                           % inserts table
        %heading
        \hline
$0$ ~~~~&~~ &$0.40825$& ~~~~&~~ $0.16667$& ~~~~~&~~ $0.25989$&~~~~~~~~~~~~\\
$0.3$ ~~~~&~~ &$0.46345$& ~~~~&~~ $0.034988$& ~~~~~&~~ $0.26889$&~~~~~~~~~~\\
$0.6$ ~~~~&~~ &$0.46433$& ~~~~&~~ $0.025041$& ~~~~~&~~ $0.26906$&~~~~~~~~~~\\
$0.9$ ~~~~&~~ &$0.46464$& ~~~~&~~ $0.020530$& ~~~~~&~~ $0.26911$&~~~~~~~~~~\\
$1.0$ ~~~~&~~ &$0.46470$& ~~~~&~~ $0.019493$& ~~~~~&~~ $0.26913$&~~~~~~~~~~\\
$1.2$ ~~~~&~~ &$0.46479$& ~~~~&~~ $0.017816$& ~~~~~&~~ $0.26914$&~~~~~~~~~~\\
$1.5$ ~~~~&~~ &$0.46488$& ~~~~&~~ $0.015955$& ~~~~~&~~ $0.26916$&~~~~~~~~~~\\
$1.8$ ~~~~&~~ &$0.46495$& ~~~~&~~ $0.014577$& ~~~~~&~~ $0.26917$&~~~~~~~~~~\\
$2.0$ ~~~~&~~ &$0.46498$& ~~~~&~~ $0.013835$& ~~~~~&~~ $0.26918$&~~~~~~~~~~\\
$2.5$ ~~~~&~~ &$0.46503$& ~~~~&~~ $0.012384$& ~~~~~&~~ $0.26919$&~~~~~~~~~~\\
$3.0$ ~~~~&~~ &$0.46507$& ~~~~&~~ $0.011310$& ~~~~~&~~ $0.26920$&~~~~~~~~~~\\
\hline
\end{tabular}
\caption{The critical values of the dimensionless event horizon radius $\tilde{r}_+^c$, magnetic charge $\tilde{Q}_m^c$, and Hawking temperature $\tilde{T}_c$ for selected values of the dimensionless nonminimal coupling parameter $\tilde{\xi}$.} % title of Table
\label{table:Fix-lamda-cri}
\end{table}
%%%%%%%%%%%
%%%%%%%%%%
%%%%%%%%%%%
%%%%%%%%%%%
\begin{figure}[ht!]
\centering
\includegraphics[width=7cm]{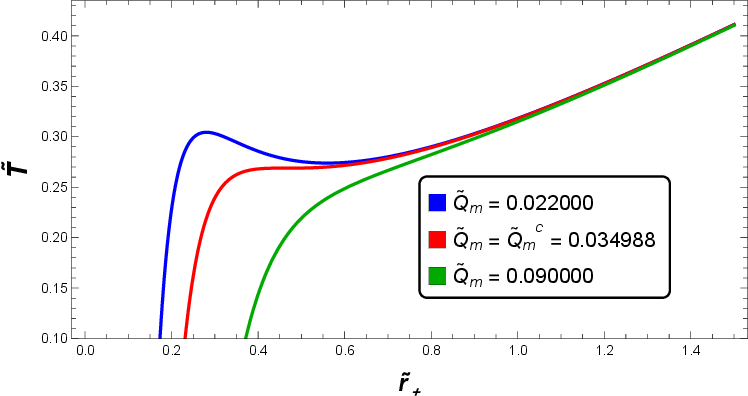}\hspace{1cm}
\includegraphics[width=7cm]{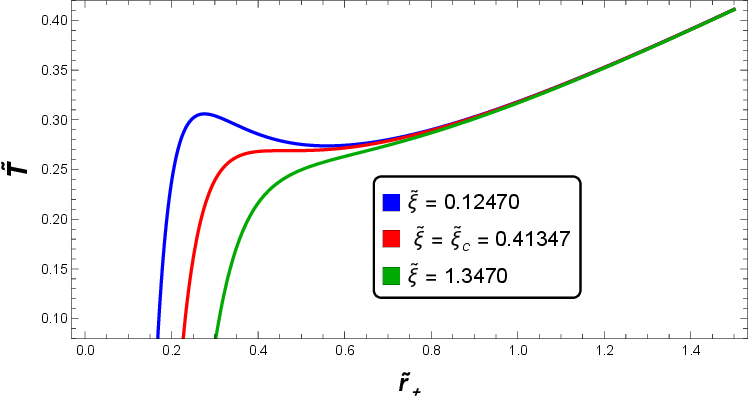}
\caption{The Hawking temperature $\tilde{T}$ as a function of the event horizon radius $\tilde{r}_+$. In the left panel, we fix $\tilde{\xi} = 0.30$, and the curves correspond to different values of the magnetic charge: $\tilde{Q}_{m} < \tilde{Q}_{m}^c$ (blue curve), $\tilde{Q}_{m} = \tilde{Q}_{m}^c$ (red curve), and $\tilde{Q}_{m} > \tilde{Q}_{m}^c$ (green curve). In the right panel, with $\tilde{Q}_m = 0.03$ fixed, the temperature is shown for different values of the nonminimal coupling parameter: $\tilde{\xi} < \tilde{\xi}_c$ (blue curve), $\tilde{\xi} = \tilde{\xi}_c$ (red curve), and $\tilde{\xi} > \tilde{\xi}_c$ (red curve). }\label{fig:rh-T}
\end{figure}
%%%%%%%%%%%
%%%%%%%%%%%
The behavior of the Hawking temperature as a function of the event horizon radius for different black hole parameters is depicted in Fig.~\ref{fig:rh-T}. In the left panel, at the critical value $\tilde{Q}_{m} = \tilde{Q}_{m}^c$ (red curve), the temperature curve develops an inflection point, signaling a second order phase transition at which the heat capacity $C_{Q}\equiv T \left(\frac{\partial S}{\partial T}\right)_{Q_m}=\left(\frac{\partial M}{\partial T}\right)_{Q_m}$ diverges.  

For the subcritical magnetic charge $\tilde{Q}_{m} < \tilde{Q}_{m}^c$ (blue curve), the temperature profile exhibits a nonmonotonic behavior, whereas
for $\tilde{Q}_{m} > \tilde{Q}_{m}^c$ (green curve), it becomes strictly monotonic, devoid of inflection points, thereby indicating the absence of phase transitions. The right panel displays a parallel phase structure under variations of the coupling parameter $\tilde{\xi}$ with fixed $\tilde{Q}_m$, reproducing the same hierarchy of thermodynamic behavior observed in the charged domain. Overall, the phase structure closely resembles that of a Van der Waals fluid \cite{Bhattacharya:2017hfj}.
%%%%%%%%%%
%%%%%%%%%%
\begin{figure}[ht!]
\centering
\subfigure[\, Typical  $\tilde{T}-\tilde{F}$ curve exhibiting a phase transition]
{\includegraphics[width=7cm]{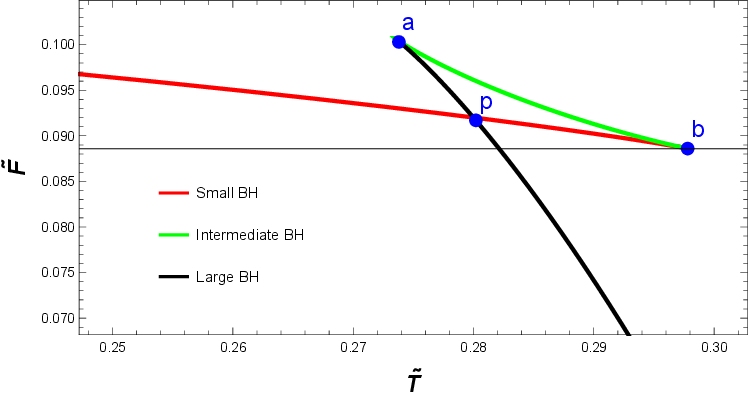}\label{fig:T-F1a}}\hspace{1cm}
\subfigure[\, Typical  $\tilde{T}-\tilde{F}$ curve without a phase transition]
{\includegraphics[width=7cm]{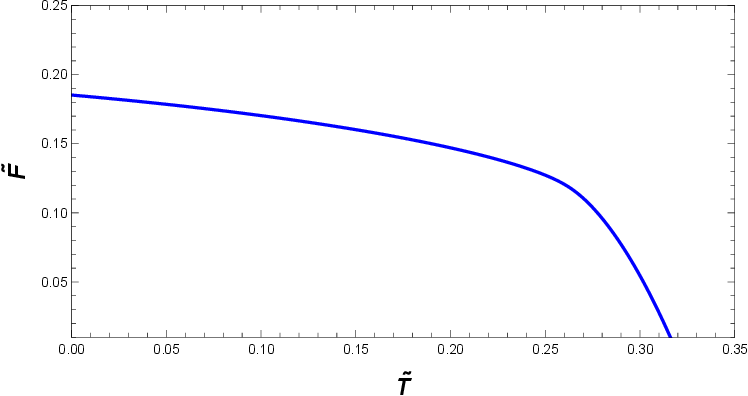}\label{fig:T-F1b}}
\caption{The typical phase transition plots show the free energy as a function of the Hawking temperature for the magnetic AdS black hole, illustrating two distinct regimes of the nonminimal coupling parameter: one exhibiting a phase transition (left panel) and the other one without a phase transition (right panel). Here, we fix the magnetic charge to $\tilde{Q}_m = 0.03$ as a representative example. In the left panel, for a fixed value $\tilde{\xi} = 0.15$ ($\tilde{\xi} < \tilde{\xi}_c$), the free energy curve exhibits three black hole branches, indicating a first order phase transition between the small and large black hole phases. In contrast, in the right panel with $\tilde{\xi} = 0.88$ ($\tilde{\xi} > \tilde{\xi}_c$), the free energy shows a single branch and no phase transition occurs, corresponding to the existence of a unique black hole solution.} \label{fig:T-Fab}
\end{figure}
%%%%%%%%%%%
%%%%%%%%%%%
\begin{figure}[ht!]
\centering
\includegraphics[width=7cm]{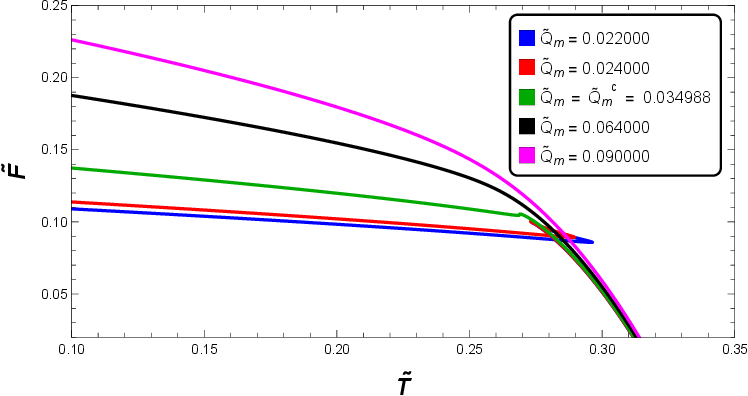}\hspace{1cm}
\includegraphics[width=7cm]{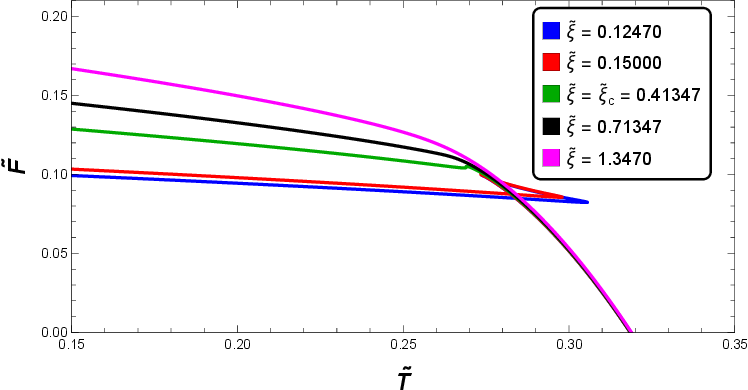}
\caption {The free energy as a function of the Hawking temperature, illustrating three distinct thermodynamic regimes. In the left panel, we fix $\tilde{\xi} = 0.30$ and vary $\tilde{Q}_m$, where the free energy curves reveal the overall trends associated with the different regimes. Similarly, in the right panel, we fix $\tilde{Q}_m = 0.03$ and vary the nonminimal coupling parameter $\tilde{\xi}$.}\label{fig:T-F34}
\end{figure}
%%%%%%%%%%%
%%%%%%%%%%%
Specifically, the typical phase structure of the free energy as a function of the Hawking temperature is displayed in Fig.~\ref{fig:T-Fab}.
For a non-monotonic $\tilde{T}(\tilde{r}_{+})$ behavior, the corresponding temperature-free energy ($\tilde{T}-\tilde{F}$) profile is shown in Fig.~\ref{fig:T-F1a}. The free energy exhibits a characteristic swallowtail structure, signaling a first order phase transition and giving rise to three distinct black hole branches, namely the small, intermediate, and large black holes. Within the temperature interval $\tilde{T}_{a} < \tilde{T} < \tilde{T}_{b}$, these branches share the same temperature but possess different free energies, indicating the coexistence and competition of multiple thermodynamic phases. A first order phase transition between the small and large black hole branches occurs at $\tilde{T}=\tilde{T}_{p}$, where their free energies coincide. The intermediate branch always has the highest free energy and is therefore thermodynamically unstable. Consequently, the small black hole phase is favored for $\tilde{T} < \tilde{T}_{p}$, while the large black hole phase becomes thermodynamically preferred for $\tilde{T} > \tilde{T}_{p}$. At the critical point, the characteristic temperatures $\tilde{T}_a$, $\tilde{T}_b$, and $\tilde{T}_p$ merge into a single inflection point at $\tilde{T}_p=\tilde{T}_c$, where the three branches coalesce into one, indicating a second order phase transition. Beyond this critical threshold, the free energy curve becomes smooth and monotonic, excluding the possibility of phase coexistence or phase transitions, as illustrated in Fig.~\ref{fig:T-F1b}. The overall phase structure behavior is further presented in Fig.~\ref{fig:T-F34}, where the magnetic charge $\tilde{Q}_m$ (or the nonminimal coupling parameter $\tilde{\xi}$) is varied while fixing $\tilde{\xi}$ (or $\tilde{Q}_m$), respectively.

So far, we have demonstrated the Van der Waals-like phase transition of the nonminimally coupled magnetic AdS black hole described by \eqref{Eq:metric}. Next we turn to the analysis of timelike geodesic motion in this black hole spacetime and show that dynamic quantities of test particles, such as the Lyapunov exponent and QPO frequencies, can also serve as effective probes of this thermodynamic phase transition.

\section{Lyapunov Exponent and Black Hole Thermodynamic Phases} \label{sec:Time-like LE}
In this section, we primarily investigate how the Lyapunov exponent of timelike particles encodes the signatures of thermodynamic phase transitions and how it evolves across different thermodynamic phases.

\subsection{Lyapunov Exponent of Timelike Circular Orbits}
The Lyapunov exponent is a fundamental quantity in the study of chaotic behavior and dynamical systems, characterizing the exponential divergence or convergence of nearby trajectories. In the context of black hole spacetimes, it provides a powerful diagnostic tool for quantifying the instability timescale of unstable geodesics in the vicinity of the event horizon. 
To proceed, we consider the Lagrangian of a test particle with mass $m$ moving in the spacetime described by the metric in \eqref{Eq:metric}, which can be written as \cite{chandrasekhar1998mathematical}
\begin{equation}
\mathcal{L}=\frac{1}{2}g_{\mu \nu}\dot{x}^{\mu}\dot{x}^{\nu}=-\frac{1}{2}N(r)\dot{t}^{2}+\frac{1}{2N(r)}\dot{r}^{2}+\frac{1}{2}{r}^{2}\dot{\theta}^{2}+\frac{1}{2}{r}^{2} \sin^{2}\theta \dot{\phi}^{2},\label{Eq:general-metric}
\end{equation}
where an overdot denotes differentiation with respect to the affine parameter $\lambda\equiv\tau/m$ ($\tau$ is the proper time). Since the black hole spacetime is static and spherically symmetric, we can restrict our analysis to geodesic motion in the equatorial plane, $\theta=\frac{\pi}{2}$.
%and the Lagrangian of test particles in this background takes the reduced form:
%\begin{equation}
%2\mathcal{L}=-N(r)\dot{t}^{2}+\frac{\dot{r}^{2}}{N(r)}+r^{2}\dot{\phi}^{2},\label{Eq:specific-metric}
%\end{equation}
The canonical momenta of the particle are given by
\begin{eqnarray}
&&p_{\mu}=\frac{\partial \mathcal{L}}{\partial \dot{x}^\mu}=\begin{cases}
p_{t}=\frac{\partial \mathcal{L}}{\partial \dot{t}}=-N(r)\dot{t}=-\mathit{E}= const.\\
p_{r}=\frac{\partial \mathcal{L}}{\partial \dot{r}}=\frac{\dot{r}}{N(r)},\\
p_{\phi}=\frac{\partial \mathcal{L}}{\partial \dot{\phi}}=r^{2}\dot{\phi} =\mathit{L}= const.
\end{cases}\label{Eq:momenta}
\end{eqnarray}
where $\mathit{E}$ and $\mathit{L}$ represent the conserved energy and angular momentum of the particle, respectively.
Solving \eqref{Eq:momenta} for the velocities yields
\begin{eqnarray}
\dot{t}=\frac{ \mathit{E}}{N(r)},&&~~~\dot{\phi}=\frac{\mathit{L}}{r^{2}}. \label{Eq:velocities}
\end{eqnarray}
To determine the radial motion of the particle, we introduce the Hamiltonian via the Legendre transformation
\cite{chandrasekhar1998mathematical,poisson2004relativist, Pugliese:2011xn}
\begin{eqnarray}
&&\mathcal{H}=p_\mu \dot{x}^\mu- \mathcal{L}=\frac{1}{2}g_{\mu \nu}\dot{x}^{\mu}\dot{x}^{\nu}=\frac{1}{2}m^2 \delta,
\label{Eq:momen}
\end{eqnarray}
where $\delta=-1$ for timelike geodesics with the normalization of the four-velocity. Substituting \eqref{Eq:general-metric}-\eqref{Eq:velocities} into \eqref{Eq:momen}, the radial equation of motion can be written as
\begin{equation}
{\dot{r}^2}+V_{\epsilon}(r)=\mathit{E}^{2},\label{Eq:radial equation}
\end{equation}
with the corresponding effective potential is given by
\begin{equation}
V_{\epsilon}(r)=N(r)\Big(m^2+\frac{L^{2}}{r^2}\Big).\label{Eq:effective-potential for timelike}
\end{equation}

The unstable circular orbit is one of the most important particle orbits around a compact object. For such orbits ($\dot{r}=0$, $\ddot{r}=0$), the effective potential satisfies the conditions \cite{Bardeen:1972fi}
\begin{equation}
V_\epsilon(r_0)=E^2~, \quad V'_{\epsilon}(r_0)=0~, \quad \text{and} \quad V''_{\epsilon}(r_0)<0~, \label{Eq:conditions}
\end{equation}
where $r_0$ is the radius of the unstable circular orbit. 

Introducing the dimensionless quantities 
\begin{equation}
    \tilde{r}_0\equiv r_0/l~, \quad \tilde{E}\equiv E/m~, \quad \tilde{L}\equiv L/(ml)~,
\end{equation}
the relations of the conserved quantities
\begin{eqnarray}
\mathit{E}^{2}=\frac{2 m^2 N^2(r)}{2 N(r)-r N'(r)}\Bigg|_{r=r_{0}},~~~~~~
\mathit{L}^{2}= \frac{m^2r^3 N'(r)}{2 N(r)-r N'(r)}\Bigg|_{r=r_{0}} 
\end{eqnarray}
can be rewritten in dimensionless form
\begin{eqnarray}
    \tilde{E}^2&=&\frac{\left(\tilde{r}_0^3 \left(-2 \tilde{M}+\tilde{r}_0^3+\tilde{r}_0\right)+\tilde{Q}_m^2 \left(2 \tilde{\xi}+\tilde{r}_0^2\right)\right)^2}{2 \tilde{Q}_m^2 \tilde{r}_0^3 \left(\tilde{M} \tilde{\xi}+(1-2 \tilde{\xi}) \tilde{r}_0^3+2 \tilde{\xi} \tilde{r}_0\right)+\tilde{r}_0^7 (\tilde{r}_0-3 \tilde{M})+4 \tilde{\xi}^2 \tilde{Q}_m^4}~,\label{relation_E}\\
    \tilde{L}^2&=&\frac{\tilde{r}_0^4 \left(\tilde{Q}_m^2 \left((6 \tilde{\xi}-1) \tilde{r}_0^4-6 \tilde{M} \tilde{\xi} \tilde{r}_0\right)+\tilde{r}_0^5 \left(\tilde{M}+\tilde{r}_0^3\right)+2 \tilde{\xi} \tilde{Q}_m^4\right)}{2 \tilde{Q}_m^2 \tilde{r}_0^3 \left(\tilde{M} \tilde{\xi}+(1-2 \tilde{\xi}) \tilde{r}_0^3+2 \tilde{\xi} \tilde{r}_0\right)+\tilde{r}_0^7 (\tilde{r}_0-3 \tilde{M})+4 \tilde{\xi}^2 \tilde{Q}_m^4}~, \label{Eq:angular momentum}
\end{eqnarray}
which are independent of $l$ and $m$.

For massless particles ($m=0$), the conditions (\ref{Eq:conditions}) for circular orbits reduce to 
\begin{eqnarray}
    &&2 \tilde{Q}_m^2 \tilde{r}_0^3 \left(\tilde{M} \tilde{\xi}+(1-2 \tilde{\xi}) \tilde{r}_0^3+2 \tilde{\xi} \tilde{r}_0\right)+\tilde{r}_0^7 (\tilde{r}_0-3 \tilde{M})+4 \tilde{\xi}^2 \tilde{Q}_m^4=0~,\\
    &&\frac{L^2}{E^2l^2}=\frac{\tilde{L}^2}{\tilde{E}^2}=\frac{2 \tilde{\xi} \tilde{Q}_m^2 \tilde{r}_0^2+\tilde{r}_0^6}{\tilde{r}_0^3 \left(-2 \tilde{M}+\tilde{r}_0^3+\tilde{r}_0\right)+\tilde{Q}_m^2 \left(2 \tilde{\xi}+\tilde{r}_0^2\right)}~.
\end{eqnarray}

The Lyapunov exponent for a massive particle moving along an unstable circular orbit is given by \cite{Cardoso:2008bp},
\begin{eqnarray}
\lambda \equiv \sqrt{\frac{-V''_{\epsilon}(r_0)}{2 \dot{t}^{2}}}=\frac{1}{2}\sqrt{\Big(r_{0}N'(r_{0})-2N(r_{0}) \Big)V''_{\epsilon}(r_{0})}~.\label{Eq:massive}
\end{eqnarray}
We introduce the dimensionless effective potential and Lyapunov exponent as
\begin{eqnarray}
    &&\tilde{V}_\epsilon\equiv \frac{V_\epsilon}{m^2}=\left(\frac{\tilde{L}^2}{\tilde{r}^2}+1\right)\frac{\tilde{r}^3 \left(-2 \tilde{M}+\tilde{r}^3+\tilde{r}\right)+\tilde{Q}_m^2 \left(2 \tilde{\xi}+\tilde{r}^2\right)}{2 \tilde{\xi} \tilde{Q}_m^2+\tilde{r}^4}~,\\
    &&\tilde{\lambda}\equiv \lambda l=\frac{1}{2}\sqrt{\Big(r_{0}N'(r_{0})-2N(r_{0}) \Big)\tilde{V}''_{\epsilon}(\tilde{r}_{0})}~,
\end{eqnarray}
where $\tilde{M}$ can be expressed in terms of $\tilde{r}_+$ using the relation (\ref{M_relation}) and
\begin{equation}
    r_{0}N'(r_{0})-2N(r_{0})=-\frac{2 \left(2 \tilde{Q}_m^2 \tilde{r}_0^3 \left(\tilde{M} \tilde{\xi}+(1-2 \tilde{\xi}) \tilde{r}_0^3+2 \tilde{\xi} \tilde{r}_0\right)+\tilde{r}_0^7 (\tilde{r}_0-3 \tilde{M})+4 \tilde{\xi}^2 \tilde{Q}_m^4\right)}{\left(2 \tilde{\xi} \tilde{Q}_m^2+\tilde{r}_0^4\right)^2}
\end{equation}
is also dimensionless.

\begin{figure}[ht!]
\centering
\includegraphics[width=7cm]{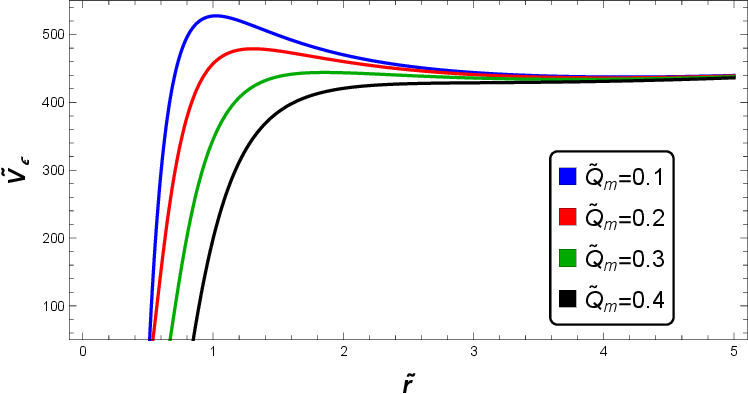}\hspace{1cm}
\includegraphics[width=7cm]{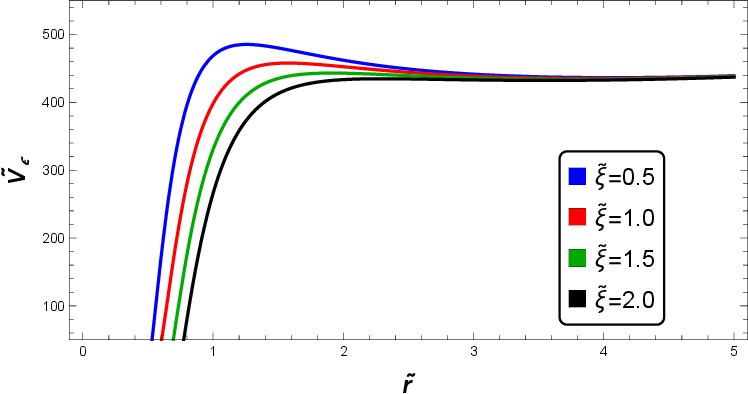}
\caption{The dimensionless effective potential $\tilde{V}_\epsilon$ as a function of the dimensionless radial coordinate $\tilde{r}$, illustrating the influence of the black hole parameters. We fix the angular momentum to $\tilde{L} = 20$ and the event horizon radius to $\tilde{r}_+=0.5$. In the left panel, the nonminimal coupling parameter is fixed as $\tilde{\xi} = 0.5$, while in the right panel the magnetic charge is fixed at $\tilde{Q}_{m} = 0.15$.}\label{fig:vf-r}
\end{figure}

In Fig.~\ref{fig:vf-r}, we illustrate how the magnetic charge $\tilde{Q}_{m}$ and the nonminimal coupling parameter $\tilde{\xi}$ influence the effective potential. In each panel, the unstable timelike circular geodesics, corresponding to the local maxima of the effective potential, gradually disappear as the parameters $\tilde{Q}_{m}$ and $\tilde{\xi}$ increase. This behavior indicates a strong connection with the Lyapunov exponent, reflecting changes in the dynamical stability of black holes across different parameter regimes.

In the following subsection, we investigate the Lyapunov exponent of unstable circular orbits in the vicinity of the magnetic AdS black hole and analyze its behavior as a dynamical probe of the thermodynamic phase transition.

\subsection{Thermodynamic Signatures in Lyapunov Exponent }

Investigating the Lyapunov exponent as a probe of the thermodynamic phase transition of the magnetic AdS black hole requires a numerical treatment, since the circular orbit conditions in Eq.~\eqref{Eq:conditions} do not admit an analytical solution for the dimensionless radius of the unstable circular orbit $\tilde{r}_0$. By imposing these constraints and fixing appropriate parameter values, we numerically determine $\tilde{r}_0$ and subsequently compute the dimensionless Lyapunov exponent $\tilde{\lambda}$. This procedure allows us to examine its dependence on the black hole parameters and to explore its behavior across different thermodynamic phases.

%%%%%%%%%%%
%%%%%%%%%%%
\begin{figure}[ht!]
\centering
\includegraphics[width=7cm]{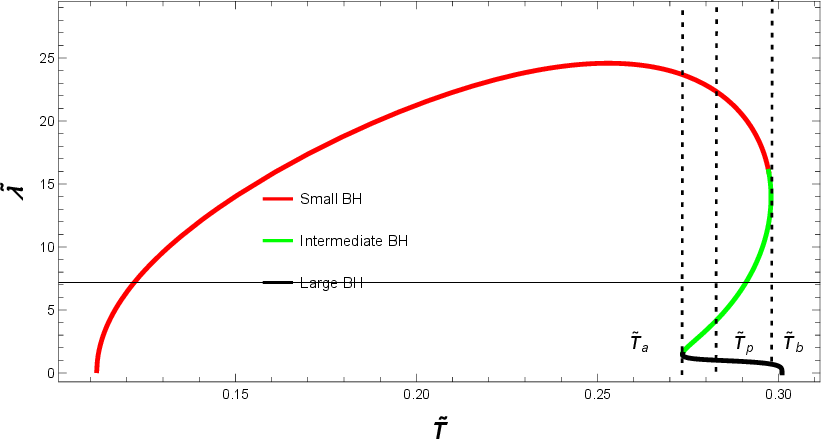}\hspace{1cm}
\includegraphics[width=7cm]{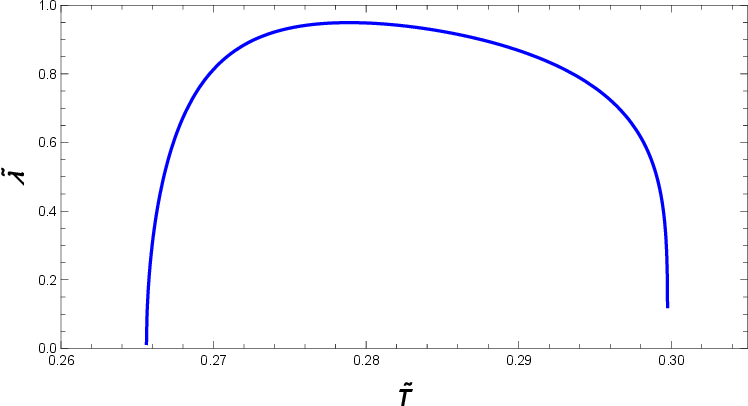}
\caption{Typical behavior of the Lyapunov exponent for a massive particle on unstable circular orbits as a function of the Hawking temperature. The angular momentum is fixed as $\tilde{L}=20$, and the black hole parameters are chosen to be the same as those in Fig.~\ref{fig:T-Fab}.}\label{fig:T-Lamda}
\end{figure}

As shown in Fig.~\ref{fig:T-Lamda}, the $\tilde{T}$–$\tilde{\lambda}$ diagram exhibits a structure closely analogous to that of the free energy profile. For $\tilde{\xi} = 0.15$ ($\tilde{\xi} < \tilde{\xi}_{c}$), the left panel displays a multivalued structure within the temperature interval $\tilde{T}_a < \tilde{T} < \tilde{T}_b$, corresponding to three distinct black hole branches.

The small black hole branch appears at lower temperatures and is associated with a larger Lyapunov exponent, indicating stronger dynamical instability. The intermediate branch lies near a threshold region and is typically unstable, while the large black hole branch emerges at higher temperatures, where the Lyapunov exponent gradually approaches zero.
In this regime, the Lyapunov exponent exhibits a characteristic branch switching behavior at the transition temperature $\tilde{T}_{p}$, reflecting the abrupt transition between the small and large black hole phases and signaling a first order phase transition. Consistent with the multivalued free energy structure for $\tilde{\xi} < \tilde{\xi}_{c}$, this behavior indicates the coexistence of distinct thermodynamic phases. 

In contrast, for $\tilde{\xi} = 0.88$ ($\tilde{\xi} > \tilde{\xi}_{c}$) in the right panel, the Lyapunov exponent shows a smooth and continuous dependence on temperature,  indicating the presence of  a single black hole phase and the absence of any phase transition. 

%%%%%%%%%%%
%%%%%%%%%%%
\begin{figure}[ht!]
\centering
\includegraphics[width=7cm]{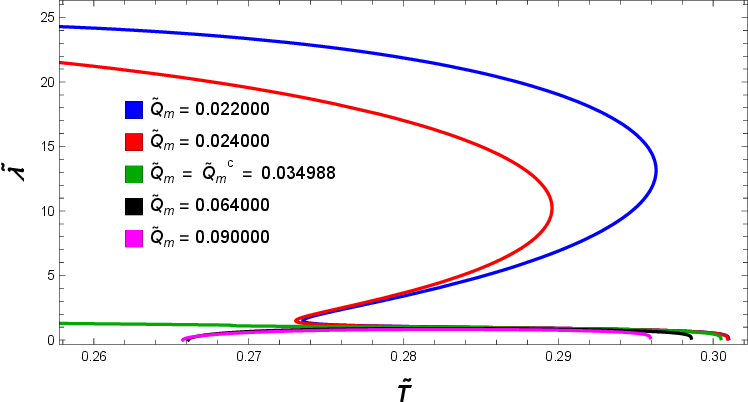}\hspace{1cm}
\includegraphics[width=7cm]{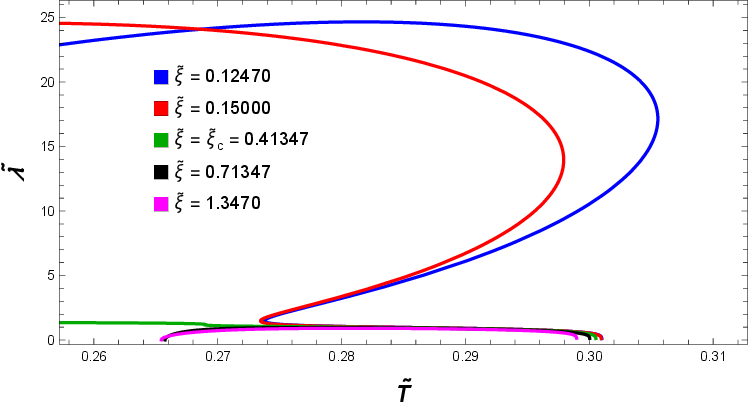}
\caption{The Lyapunov exponent of unstable circular orbits as a function of the Hawking temperature. In each panel, the Lyapunov exponent exhibits distinct behaviors in the presence and absence of a thermodynamic phase transition. The angular momentum is fixed as $\tilde{L}=20$, and the black hole parameters are chosen to be the same as those in Fig.~\ref{fig:T-F34}.}\label{fig:T-Lamda-overall}
\end{figure}
%%%%%%%%%%
%%%%%%%%%%
The overall phase transition behavior of the Lyapunov exponent as a function of the Hawking temperature is summarized in Fig.~\ref{fig:T-Lamda-overall}. The plots reveal a coherent pattern across different black hole parameter regimes, clearly distinguishing first order, second order and no-transition behaviors within a unified framework. This highlights the Lyapunov exponent as an effective dynamical probe of the underlying thermodynamic phase structure.

In the subcritical regime (e.g., $\tilde{Q}_m < \tilde{Q}_{m}^c$ or $\tilde{\xi} < \tilde{\xi}_c$), the Lyapunov exponent exhibits a multivalued structure (blue and red curves), reflecting the coexistence of multiple black hole branches and signaling a first order phase transition.

At the critical point (e.g., $\tilde{Q}_m = \tilde{Q}_{m}^c$ or $\tilde{\xi} = \tilde{\xi}_c$), the multibranch structure collapses into a single-valued curve (green curve). The closure of the Lyapunov exponent gap provides a clear dynamical signature of the second order phase transition.

In the supercritical regime (e.g., $\tilde{Q}_m > \tilde{Q}_{m}^c$ or $\tilde{\xi} > \tilde{\xi}_c$), the Lyapunov exponent remains single-valued (black and magenta curves), corresponding to a unique black hole phase without thermodynamic competition.

Notably, unlike the $\tilde{T}-\tilde{F}$ curves, the $\tilde{T}-\tilde{\lambda}$ curves remain smooth across the entire temperature range. This distinction originates from the fundamentally different roles of the two quantities: the free energy governs global thermodynamic stability and phase selection, whereas the Lyapunov exponent characterizes local dynamical instability of particle motion.
Overall, the Lyapunov exponent establishes a direct connection between dynamical instability of particle orbits and black hole thermodynamics, providing a complementary dynamical diagnostic for identifying and characterizing black hole phase transitions.

%%%%%%%%%%%
\section{Quasi-Periodic Oscillations and Black Hole Thermodynamic Phases}
\label{sec:QPO phase transition}
In this section, we investigate how the QPO frequencies of test particles encode the thermodynamic phase structure of the nonminimally coupled magnetic AdS black hole. In particular, we analyze their characteristic behavior across different phase transition regimes.

\subsection{QPO Frequencies of Timelike Circular Orbits}

To analyze the fundamental frequencies associated with the oscillatory motion of a test particle around the nonminimally coupled magnetic black hole, we consider small perturbations around a circular orbit 
\begin{equation}
    r \to r_{0}+\delta r~, \quad \theta \to \theta_0+\delta \theta~.
\end{equation}

Following \cite{Xamidov:2025hrj}, we introduce an effective potential $H_{\rm pot}$ from (\ref{Eq:momen}),
\begin{equation}
    \frac{1}{2}g_{rr}\dot{r}^2+\frac{1}{2}g_{\theta\theta}\dot{\theta}^2=-m^2 H_{\rm pot}~, \label{H_pot}
\end{equation}
where 
\begin{equation}
    H_{\rm pot}=\frac{1}{2}+\frac{1}{2m^2}g_{tt}\dot{t}^2+\frac{1}{2m^2}g_{\phi\phi}\dot{\phi}^2=\frac{1}{2}-\frac{E^2}{2m^2N(r)}+\frac{L^2}{2m^2 r^2\sin^2{\theta}}~.
\end{equation}

Its relation to a general potential $V_\epsilon\left(r,\theta\right)$ is
\begin{equation}
    H_{\rm pot}=\frac{1}{2m^2N(r)}\left(V_\epsilon\left(r,\theta\right)-E^2\right)~.
\end{equation}

Expanding $H_{\rm pot}\left(r,\theta\right)$ around the circular orbit $\left(r_0,\theta_0\right)$, we obtain
\begin{eqnarray}
    H_{\rm pot}\left(r,\theta\right)&=&H_{\rm pot}\left(r_0,\theta_0\right)+\delta r \partial_r H_{\rm pot}\left(r,\theta\right) \big|_{r_0,\theta_0}+\delta \theta \partial_\theta H_{\rm pot}\left(r,\theta\right) \big|_{r_0,\theta_0}+\delta r \delta \theta \partial_r \partial_\theta H_{\rm pot}\left(r,\theta\right) \big|_{r_0,\theta_0} \notag\\
    &&+\frac{1}{2}\delta r^2 \partial_r^2 H_{\rm pot}\left(r,\theta\right) \big|_{r_0,\theta_0}+\frac{1}{2}\delta \theta^2 \partial_\theta^2 H_{\rm pot}\left(r,\theta\right) \big|_{r_0,\theta_0}+\mathcal{O}\left(\delta ^3\right)~,
\end{eqnarray}
where the first four terms vanish for equatorial orbits $\theta_0=\frac{\pi}{2}$ and via (\ref{H_pot}) we have
\begin{equation}
    g_{rr}\delta \dot{r}^2 +g_{\theta\theta}\delta \dot{\theta}^2=-m^2\delta r^2 \partial_r^2 H_{\rm pot}\left(r,\theta\right) \big|_{r_0,\theta_0}-m^2\delta \theta^2 \partial_\theta^2 H_{\rm pot}\left(r,\theta\right) \big|_{r_0,\theta_0}~.
\end{equation}
Taking the derivative with respect to the affine parameter $\lambda$, we obtain 
\begin{equation}
    \delta \dot{r}\left[g_{rr}\delta \ddot{r}+ m^2\partial_r^2 H_{\rm pot}\left(r,\theta\right) \big|_{r_0,\theta_0}\delta r\right]+\delta \dot{\theta}\left[g_{\theta\theta}\delta \ddot{\theta}+m^2 \partial_\theta^2 H_{\rm pot}\left(r,\theta\right) \big|_{r_0,\theta_0}\delta \theta\right]=0~,
\end{equation}
which leads to the decoupled harmonic oscillator equations
\begin{equation}
   \frac{d^2\delta r}{d\lambda^2}+\frac{m^2}{g_{rr}}\partial_r^2 H_{\rm pot}\left(r,\theta\right) \big|_{r_0,\theta_0}\delta r=0~, \quad \frac{d^2\delta \theta}{d\lambda^2}+\frac{m^2}{g_{\theta\theta}}\partial_\theta^2 H_{\rm pot}\left(r,\theta\right) \big|_{r_0,\theta_0}\delta \theta=0~.
\end{equation}
Transforming to coordinate time $t$, we obtain
\begin{equation}
   \frac{d^2\delta r}{dt^2}+\Omega_r^2\delta r=0~, \quad \frac{d^2\delta \theta}{dt^2}+\Omega_\theta^2\delta \theta=0~,
\end{equation}
with the characteristic frequencies
\begin{eqnarray}
    \Omega_r^2 &\equiv& \frac{m^2 N(r)^2}{E^2g_{rr}}\partial_r^2 H_{\rm pot}\left(r,\theta\right) \big|_{r_0,\frac{\pi}{2}}=\frac{N(r_0)^2}{2E^2}V_\epsilon''\left(r_0\right) ~\\
    \Omega_\theta^2 &\equiv& \frac{m^2 N(r)^2}{E^2g_{\theta\theta}}\partial_\theta^2 H_{\rm pot}\left(r,\theta\right) \big|_{r_0,\frac{\pi}{2}}=\frac{N(r_0)^2 L^2}{E^2 r_0^4}~,
\end{eqnarray}
which shows that unstable circular orbits correspond to $\Omega_r^2<0$, 
implying an imaginary radial frequency $\Omega_r$ and hence an exponential instability rather than oscillatory motion. Finally the azimuthal (Keplerian) frequency is
\begin{equation}
    \Omega_\phi\equiv\frac{d\phi}{dt}=\frac{\dot{\phi}}{\dot{t}}=\frac{LN(r_0)}{Er_0^2}=\sqrt{\frac{N'(r_0)}{2r_0}}=\Omega_\theta~,
\end{equation}
indicating the degeneracy between the vertical and orbital frequencies in spherically symmetric spacetimes.

Usually in asymptotically flat spacetimes, it is customary to express the orbital frequencies in physical units of Hertz (Hz) through
\begin{equation}
\nu_{d}= \frac{c^{3}}{2 \pi G M}\tilde{\Omega}_d \simeq 3.233\times 10^4\left(\frac{M_\odot}{M_{\rm BH}}\right)\tilde{\Omega}_d~{\rm Hz}~, \quad \tilde{\Omega}_d\equiv M~\Omega_d~ \label{Eq:Hz}
\end{equation}
where ${\rm d}=\left(r, \theta, \phi\right)$.
In the present work, however, all quantities are rendered dimensionless using the AdS radius $l$. It is therefore more appropriate to introduce $\tilde{\Omega}_d'\equiv l~\Omega_d$, rather than $\tilde{\Omega}_d\equiv M~\Omega_d$, since the black hole mass is not fixed along the curves considered here. Accordingly,
\begin{equation}
\frac{2l}{r_s}\nu_{d}=\frac{c^{3}}{2 \pi G M_\odot}\tilde{\Omega}_d' \simeq 3.233\times 10^4~\tilde{\Omega}_d'~{\rm Hz}~, \label{Eq:Hz2}
\end{equation}
where $r_s \equiv 2G M_\odot/c^2$ is the Schwarzschild radius corresponding to one solar mass.  Then using Eqs.\eqref{Eq:mass} and \eqref{Eq:angular momentum}, one can determine the stable orbit radius $\tilde{r}_0$ for a given event horizon radius $\tilde{r}_+$ with fixed $\tilde{L}$, $\tilde{Q}_m$, and $\tilde{\xi}$. The corresponding QPO frequencies, characterized by $\tilde{\Omega}_d'$ or equivalently $\frac{2l}{r_s}\nu_d$, can then be obtained, yielding the relation between the horizon radius and the QPO frequencies.

%%%%%%%%%%%
%%%%%%%%%%%
\begin{figure}[ht!]
\centering
\includegraphics[width=7cm]{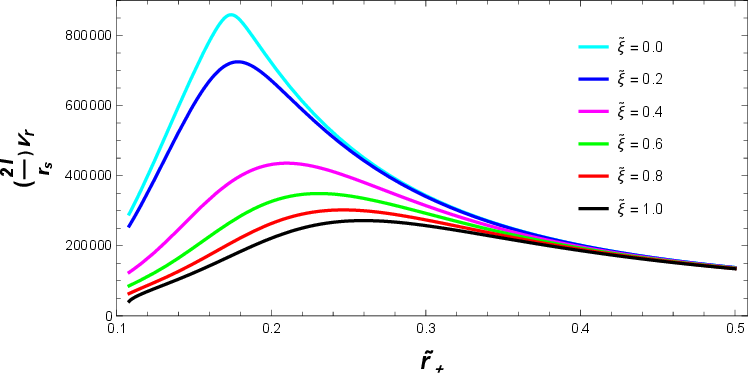}\hspace{1cm}
\includegraphics[width=7cm]{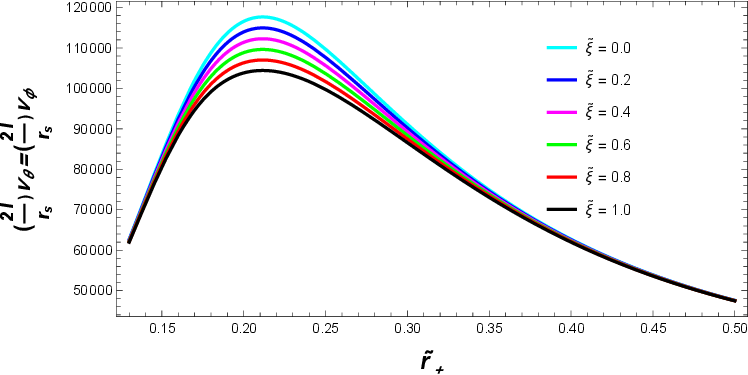}
\caption{Typical plots of the radial $\left(\frac{2l}{r_s}\right)\nu_r$ (left panel) and vertical $\left(\frac{2l}{r_s}\right)\nu_\theta = \left(\frac{2l}{r_s}\right)\nu_\phi$ (right panel) oscillation frequencies of stable circular orbits as functions of the event horizon radius $\tilde{r}_+$ for different values of the coupling parameter $\tilde{\xi}$. Here, the angular momentum of the test particle and the black hole magnetic charge are fixed at $\tilde{L}=20$ and $\tilde{Q}_m=0.03$, respectively.}\label{fig:r-Frequency}
\end{figure}

Fig.~\ref{fig:r-Frequency} shows the radial $\left(\frac{2l}{r_s}\right)\nu_r$ and vertical $\left(\frac{2l}{r_s}\right)\nu_\theta=\left(\frac{2l}{r_s}\right)\nu_\phi$ oscillation frequencies as functions of the event horizon radius $\tilde{r}_+$ for different values of the coupling parameter $\tilde{\xi}$. Both the radial and orbital frequencies exhibit a rise-peak-decline behavior as the horizon radius $\tilde{r}_{+}$ increases. The results highlight the significant role of the nonminimal coupling parameter $\tilde{\xi}$ in suppressing the oscillation frequencies, compared with the RN-AdS case ($\tilde{\xi}=0$). In particular, for black hole radii around roughly $\tilde r_{+}\sim 0.2$, i.e. $r_+ \sim 0.2 l$, the effects of $\tilde{\xi}$ become more pronounced. Moreover, each curve exhibits a peak near this region, indicating that when the black hole radius is about one fifth of the AdS radius, the oscillatory responses are strongest and the perturbed particle returns most efficiently to the stable circular orbit (equilibrium position).

\subsection{Thermodynamic Signatures in QPO Spectra}

To proceed, we investigate the twin-peak QPOs in the present background. The characteristic frequencies of the modeled QPOs, particularly the upper and lower branches, are determined within established theoretical frameworks~\cite{Shahzadi:2021upd}. We begin with the RP model, in which the upper and lower frequencies are identified with the azimuthal and periastron precession frequencies, denoted by $\nu_{U}$ and $\nu_{L}$, respectively. In RP model, the upper and lower QPO frequencies are given by~\cite{stella1999correlations}
\begin{equation}
\nu_{U}= \nu_{\phi}, \qquad 
\nu_{L}=\nu_{\phi}-\nu_{r}.
\label{Eq:RP model}
\end{equation}
We then analyze the behavior of these frequencies as functions of the black hole temperature and examine their behavior across the thermodynamic phase transitions of the nonminimally coupled magnetic black holes. 
%We present a graphical analysis demonstrating that phase transitions in black hole thermodynamics can be effectively traced through the upper and lower QPO frequencies. 

%%%%%%%%%%%
%%%%%%%%%%%
\begin{figure}[ht!]
\centering
\subfigure[\, Typical  $\tilde{T}$-Upper/Lower frequencies profile with phase transition]
{\includegraphics[width=7cm]{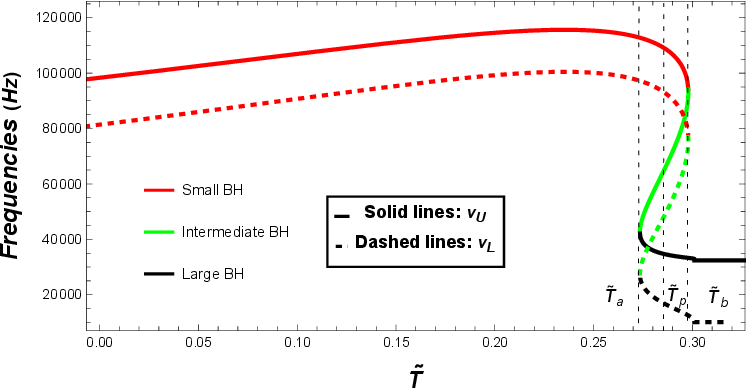}\label{fig:T-U}}\hspace{1cm}
\subfigure[\, Typical  $\tilde{T}$-Upper/Lower frequencies profile without phase transition]
{\includegraphics[width=7cm]{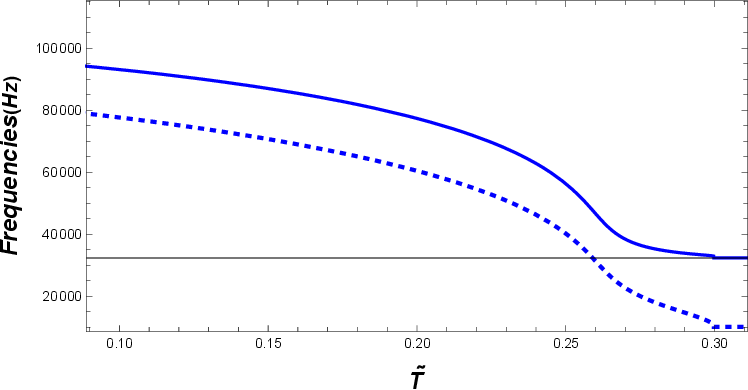}\label{fig:T-U1}}
\caption{Typical behavior of the upper $\left(\frac{2l}{r_s}\right)\nu_U$ and lower $\left(\frac{2l}{r_s}\right)\nu_L$ QPO frequencies as functions of the Hawking temperature, illustrating the cases with (left) and without (right) thermodynamic phase transitions. For simplicity, the overall factor $\left(\frac{2l}{r_s}\right)$ is omitted in the plots. The black hole parameters are the same as those used in Fig.~\ref{fig:T-Fab} and Fig.~\ref{fig:T-Lamda}. The left panel shows the existence of three black hole branches and demonstrates the first order phase transition between the small  and large black hole phases. By contrast, the right panel exhibits a single smooth branch, indicating the absence of any phase transition.}\label{fig:T-U2}
\end{figure} 
%%%%%%%%%%%

Fig.~\ref{fig:T-U} illustrates the behavior of the upper (solid curves) and lower (dashed curves) QPO frequencies as functions of the black hole Hawking temperature for $\tilde{\xi} = 0.15~\left(\tilde{\xi} < \tilde{\xi}_c\right)$. The frequency curves clearly reveal the thermodynamic phase structure of the black hole, exhibiting multiple branches. This behavior closely mirrors the thermodynamic features observed earlier in the analyses of the free energy and Lyapunov exponent. Such a consistent correspondence between the black hole phase transition and the QPO spectrum suggests a promising avenue for both theoretical investigation and potential observational verification.

Along the small black hole branch, the upper and lower QPO frequencies exhibit a nonmonotonic dependence on the Hawking temperature: they initially increase with $\tilde{T}$, reach a maximum, and then decrease as the branch approaches the phase transition point. Along the intermediate black hole branch, both frequencies increase monotonically with $\tilde{T}$, whereas on the large black hole branch, they display a monotonic decrease, reflecting the gradual suppression of the QPO spectrum as the system evolves toward the large black hole phase.

It is worth noting that, for $\tilde{T}<\tilde{T}_p$, the small black hole branch is thermodynamically stable, while for $\tilde{T}>\tilde{T}_p$, the large black hole branch becomes stable. Therefore, at the phase transition temperature $\tilde{T}=\tilde{T}_p$, the QPO frequencies undergo a discontinuous jump from the small black hole branch to the large black hole branch. In contrast, when the nonminimal coupling parameter exceeds its critical value, e.g., $\tilde{\xi} = 0.88~(\tilde{\xi} > \tilde{\xi}_c)$, the upper and lower frequencies evolve smoothly and monotonically with temperature, as shown in Fig.~\ref{fig:T-U1}. This behavior indicates the existence of a single thermodynamically stable phase without any phase transition. Such a trend appears to be universal in this spacetime background, reflecting the generic thermodynamic behavior of AdS black holes through the perspective of QPO frequencies.

%%%%%%%%%%%
\begin{figure}[ht!]
\centering
\includegraphics[width=7cm]{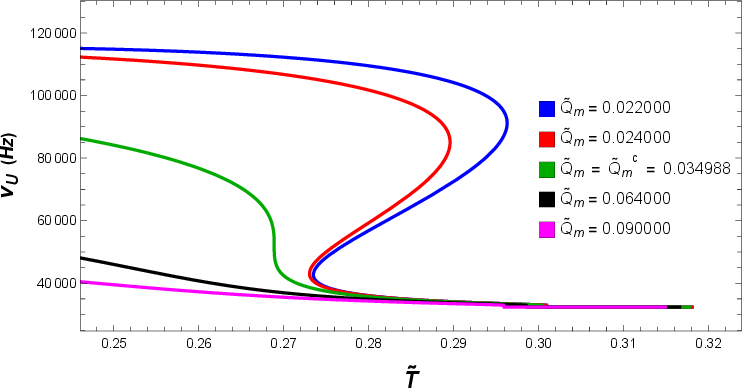}\hspace{1cm}
\includegraphics[width=7cm]{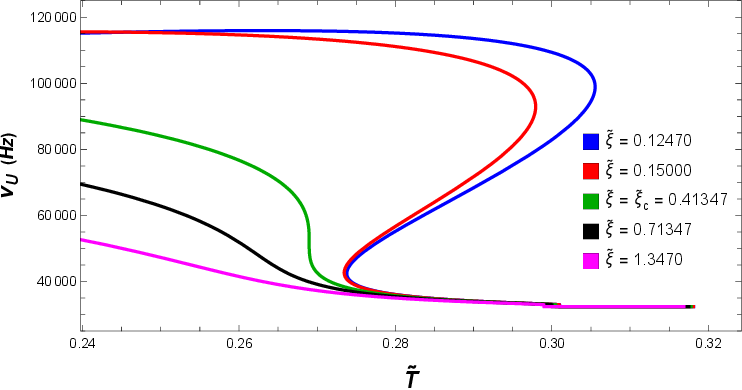}
\includegraphics[width=7cm]{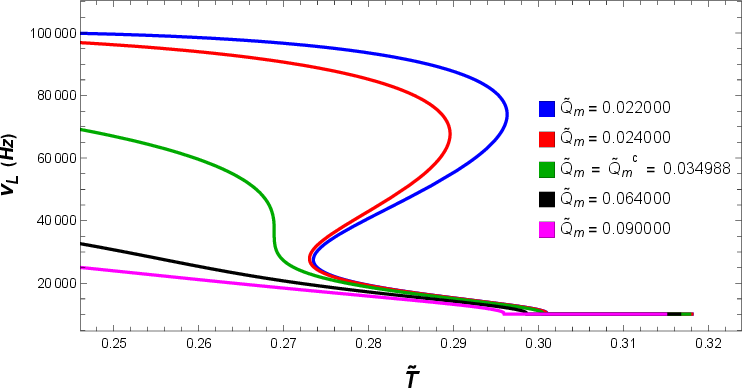}\hspace{1cm}
\includegraphics[width=7cm]{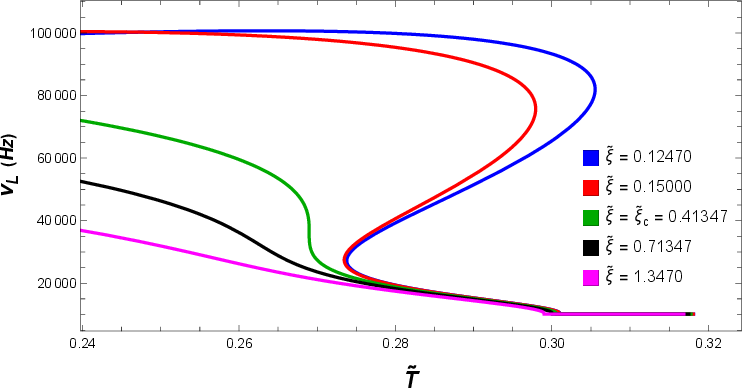}
\caption{The upper and lower QPO frequencies as functions of the Hawking temperature, illustrating how the QPO spectrum evolves with temperature and revealing the presence or absence of thermodynamic phase transitions in the system. The black hole parameters are the same as those used in Fig.~\ref{fig:T-F34} and Fig.~\ref{fig:T-Lamda-overall}.}\label{fig:T-UL-overall}
\end{figure}
%%%%%%%%%%%

The unified thermodynamic behavior of the upper and lower QPO frequencies as functions of the Hawking temperature is summarized in Fig.~\ref{fig:T-UL-overall}. This comprehensive analysis identifies the first order, second order, and supercritical regimes within a single dynamical framework, analogous to the Lyapunov exponent approach for diagnosing black hole phase transitions. In the subcritical regime, the upper and lower frequencies exhibit a multivalued structure, signaling a first order phase transition and the coexistence of multiple black hole phases. At the critical point, the three black hole branches merge into a smooth continuous curve, indicating the onset of a second order phase transition. In the supercritical regime, the frequencies become single-valued, corresponding to a unique thermodynamically stable phase.

The apparent breakpoints appearing near $\tilde T\sim 0.3$ are associated with changes in the structure of admissible stable circular orbits for fixed angular momentum. In this regime, the corresponding stable geodesic branch approaches the boundary of orbital stability, leading to rapid variations or termination of the QPO solutions.

To examine whether the thermodynamic signatures identified in the main text are robust against different QPO prescriptions, we further analyze several well established QPO models. The corresponding defination and results are presented in Appendix~\ref{Thermodynamic Imprints in Different QPO Models}.

Hereby, we establish a clear and quantitative correlation between black hole thermodynamic phase transitions and the QPO frequencies of orbiting particles, exhibiting behavior closely analogous to the thermodynamic signatures encoded in the Lyapunov exponent. These results suggest that QPOs can serve as an effective dynamical probe of black hole thermodynamic phase transitions, providing a useful framework for exploring the interplay among strong gravity, orbital dynamics, and black hole thermodynamics.

\section{Mapping QPO–Lyapunov Exponent Across Thermodynamic Phases}
\label{Mapping QPO–Lyapunov Exponent Across Thermodynamic Phases}
It is worth emphasizing that the Lyapunov exponent and the QPO frequencies originate from different classes of circular orbits. The Lyapunov exponent,
\begin{equation}
    \lambda^2=-\frac{N(r_0)^2 V_\epsilon''(r_0)}{2E^2}
    =N'(r_0)^2-\frac{1}{2} N(r_0) \left(N''(r_0)+\frac{3 N'(r_0)}{r_0}\right)
\end{equation}
is evaluated at the radius $r_0$ of the unstable circular orbit and characterizes the instability timescale of radial perturbations. In contrast, the QPO frequencies are associated with stable circular orbits. In particular, the radial epicyclic frequency is given by
\begin{equation}
    \Omega_r^2=\frac{N(r_0)^2 V''(r_0)}{2 E^2}
    =\frac{1}{2} N(r_0) \left(N''(r_0)+\frac{3 N'(r_0)}{r_0}\right)-N'(r_0)^2,
\end{equation}
where $r_0$ now denotes the radius of the stable circular orbit. One immediately observes that the two expressions differ only by an overall sign, reflecting the fact that the Lyapunov exponent describes orbital instability, whereas the epicyclic frequency characterizes stable oscillatory motion around equilibrium.

For fixed black hole parameters $\tilde{M}$, $\tilde{Q}_m$, and $\tilde{\xi}$, a given unstable orbit radius $\tilde{r}_0$ uniquely determines the dimensionless Lyapunov exponent $\tilde{\lambda}$. Similarly, a given stable orbit radius $\tilde{r}_0$ determines the dimensionless radial oscillation frequency $\tilde{\Omega}_d$. Nevertheless, the unstable and stable circular orbit radii are, in general, independent variables, and therefore no direct functional relation between the Lyapunov exponent and the QPO frequencies can be established at this stage. A nontrivial connection emerges only after fixing the parameters $\tilde{L}$, $\tilde{Q}_m$, and $\tilde{\xi}$. In this case, a given horizon radius $\tilde{r}_+$ simultaneously determines both the unstable and stable circular orbit radii, thereby fixing the Lyapunov exponent and the QPO frequencies at the same time. Consequently, both quantities become indirectly correlated through their common dependence on the black hole horizon radius and thermodynamic background.

\begin{figure}[ht!]
\centering
\subfigure[\, Regular magnetic AdS black hole with thermodynamic phase transitions]
{\includegraphics[width=7cm]{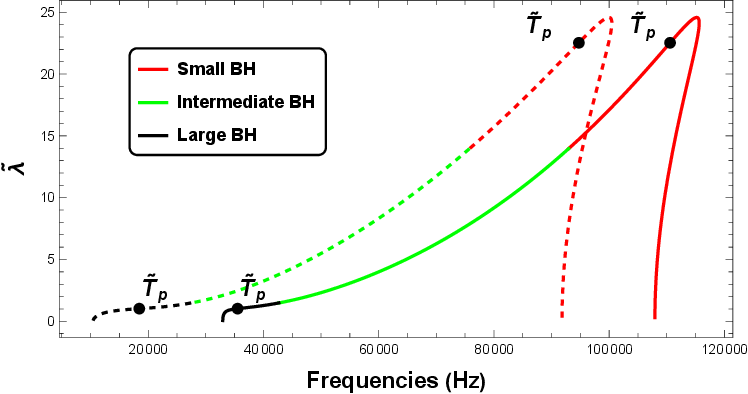}\label{fig:Q1}}\hspace{1cm}
\subfigure[\, Regular magnetic AdS black hole without thermodynamic phase transitions]
{\includegraphics[width=7cm]{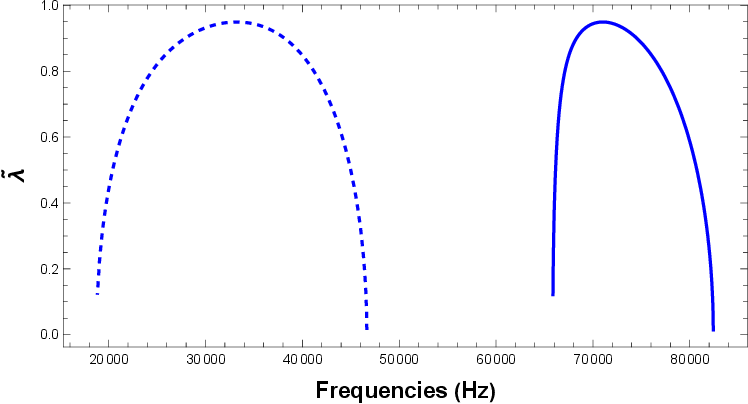}\label{fig:Q2}}
\subfigure[\, RN-AdS black hole with thermodynamic phase transitions]
{\includegraphics[width=7cm]{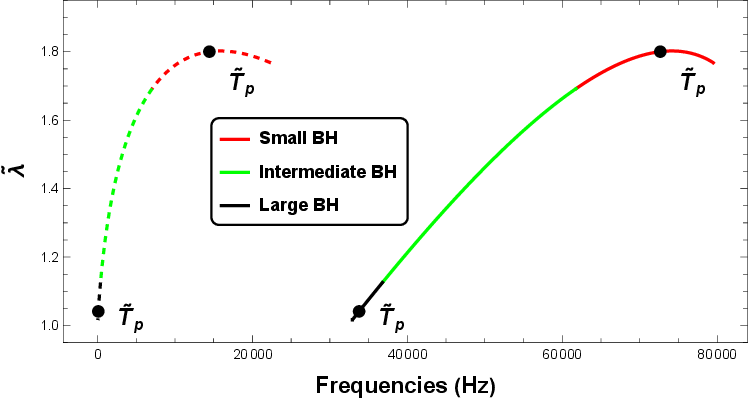}\label{fig:Q3}}\hspace{1cm}
\subfigure[\, RN-AdS black hole without thermodynamic phase transitions]
{\includegraphics[width=7cm]{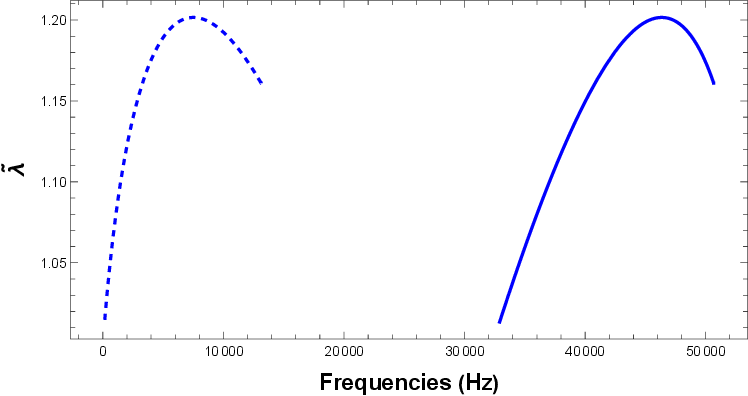}\label{fig:Q4}}
\subfigure[\, Asymptotically flat RN black hole without thermodynamic phase transitions]
{\includegraphics[width=8cm]{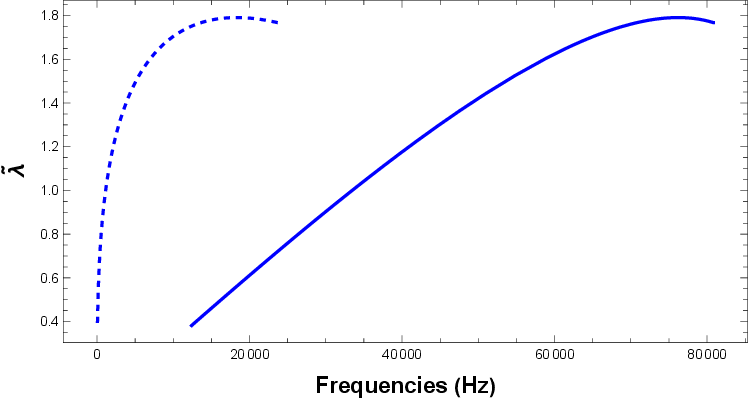}\label{fig:Q5}}\hspace{1cm}
\caption{Representative $\tilde{\lambda}$–QPO frequencies profiles for different black hole spacetimes. The upper and lower QPO frequencies are denoted by $\nu_U$ (solid curves) and $\nu_L$ (dashed curves), respectively. Panels (a) and (c) correspond to black holes exhibiting thermodynamic phase transitions, with the same parameters as those adopted in Fig.~\ref{fig:T-Fab}, Fig.~\ref{fig:T-Lamda} and Fig.~\ref{fig:T-U2}, where three distinct branches associated with small, intermediate, and large black hole phases emerge. Panels (b), (d), and (e) represent cases without phase transitions. These results indicate that the mapping between QPO frequencies and the Lyapunov exponent retains clear signatures of the underlying thermodynamic phase structure.}\label{fig:Q-overall}
\end{figure}

Since the Hawking temperature is not directly observable, establishing a connection between the Lyapunov exponent and QPO frequencies is of particular importance. As illustrated in Figs.~\ref{fig:T-Lamda} and \ref{fig:T-U2}, both the Lyapunov exponent and the QPO frequencies are single-valued functions of the Hawking temperature along each thermodynamic branch, regardless of whether phase transitions occur. This allows the Hawking temperature to be eliminated as a parameter, leading to a direct branch dependent mapping between the Lyapunov exponent and the observable QPO frequencies, as shown in Fig. \ref{fig:Q-overall}. A qualitatively similar QPO–Lyapunov exponent correspondence is observed in the magnetic AdS, RN–AdS, and RN black holes, indicating that the relation is not restricted to a particular black hole solution.

%%%%%%%%%%%

From Fig.~\ref{fig:Q1}, the relation between the Lyapunov exponent and the QPO frequencies can be obtained for each of the small, intermediate, and large black hole branches, although the segment connecting the two states at $T_p$ on the small and large branches is thermodynamically unstable. Note that, for the small black hole phase, one $\tilde{\lambda}$ may correspond to two values of the upper/lower QPO frequency. Interestingly, within a small range of QPO frequencies, one upper/lower frequency may also correspond to two values of $\tilde{\lambda}$.
On the other hand, in Fig.~\ref{fig:Q2}, which corresponds to the case without phase transitions, one frequency corresponds to only one value of $\tilde{\lambda}$, while one $\tilde{\lambda}$ may still correspond to two values of the upper/lower QPO frequency. 

Panels (c) and (d) exhibit similar behaviors for the Reissner-Nordström (RN)-AdS black hole, indicating that this relation is generic for AdS black holes. More importantly, for the asymptotically flat RN black hole without such Van der Waals-like phase transitions, in Fig.~\ref{fig:Q5}, a similar relation between the Lyapunov exponent and the QPO frequencies still holds. This indicates that the Lyapunov exponent of unstable circular orbits can be inferred from the QPO frequencies of stable circular orbits for a fixed angular momentum (or energy, or the ratio of angular momentum to energy) of a particle.

Unlike the eikonal QNM correspondence, where both the oscillation frequency and Lyapunov exponent are governed by the same unstable null circular orbit~\cite{Cardoso:2008bp}, the present relation between QPO frequencies and the Lyapunov exponent is indirect. In our case, the QPO frequencies originate from stable timelike circular motion, whereas the Lyapunov exponent characterizes unstable circular orbits. The observed correlation emerges because both quantities are simultaneously determined by the same black hole thermodynamic background once the conserved particle angular momentum is fixed. For a family of particles with the same conserved angular momentum, the black hole spacetime uniquely correlates the instability timescale of unstable circular orbits with the QPO frequencies of stable circular orbits.

Overall, the QPO frequency–Lyapunov exponent mapping retains clear imprints of black hole thermodynamic phases. The mapping inherits the same branch structure, preserving the thermodynamic classification shown in Figs.~\ref{fig:T-Lamda} and \ref{fig:T-U2}. Since both the QPO frequencies and the Lyapunov exponent are determined by the underlying black hole spacetime geometry, their correspondence survives even in the absence of phase transitions. This relation provides a potential observational avenue for identifying chaotic orbital behavior and probing black hole phase structures through QPO measurements.

%%%%%%%%%%%
%%%%%%%%%%%
\section{Conclusion}
\label{sec:conclusion}
In this work, we investigated the interplay between black hole thermodynamics and timelike particle dynamics in a nonminimally coupled magnetic AdS black hole. The free energy analysis revealed a Van der Waals-like phase structure consisting of small, intermediate, and large black hole branches, together with first  and second order phase transitions.

We first analyzed unstable timelike circular geodesics and evaluated the corresponding Lyapunov exponent. The resulting temperature–Lyapunov exponent relation exhibits a branch structure closely analogous to that of the thermodynamic free energy. In the subcritical regime, the Lyapunov exponent becomes multivalued, reflecting the coexistence of different black hole phases, while at the critical point the multiple branches merge into a single curve. These results demonstrate that orbital instability provides a useful dynamical probe of black hole's thermodynamical phase transitions.

We then studied QPO frequencies associated with perturbations of stable circular orbits within the RP model. The upper and lower QPO frequencies exhibit thermodynamic branch structures analogous to those found in the Lyapunov exponent. The corresponding temperature-frequency relations clearly distinguish first order, second order, and supercritical regimes, indicating that QPO observables can effectively encode information about the underlying thermodynamic phase structure.

The most important result of this work is the establishment of a mapping between QPO frequencies and the Lyapunov exponent. Unlike the well known eikonal correspondence between quasinormal modes and unstable null geodesics \cite{Cardoso:2008bp}, the present relation connects quantities originating from stable and unstable timelike circular orbits. We show that both quantities are controlled by the same black hole spacetime geometry and therefore exhibit a nontrivial correlation. Remarkably, the QPO–Lyapunov exponent mapping inherits the thermodynamic branch structure and remains qualitatively valid even when thermodynamic phase transitions are absent. Furthermore, the same qualitative QPO–Lyapunov exponent correspondence is observed in the nonminimal magnetic AdS, RN–AdS, and RN black holes, suggesting that it may represent a generic feature of black hole spacetimes rather than a property of any particular solution. Remarkably,  we found  that this correspondence survives even in the absence of thermodynamic phase transitions, hinting at a deeper connection between orbital instability and oscillatory dynamics. To the best of our knowledge, such a QPO–Lyapunov exponent relation has not been explored previously.

Our results reveal a new link among orbital instability, QPO phenomenology, and black hole thermodynamics. They suggest that observationally accessible QPO signals may provide indirect information about chaotic orbital behavior and the thermodynamic state of black holes. This opens a new avenue for connecting black hole dynamics, chaos, and thermodynamics within a unified framework.

\section*{Acknowledgments}
This work was supported in part by the National Natural Science Foundation of China under Grant No. 12375054. Z.-Y. Tang was supported by the Institute for Basic Science (IBS) under Project Code IBS-R018-D3.

\appendix

\section{Thermodynamic Imprints in Different QPO Models}
\label{Thermodynamic Imprints in Different QPO Models}

\begin{figure}[ht!]
\centering
\includegraphics[width=7cm]{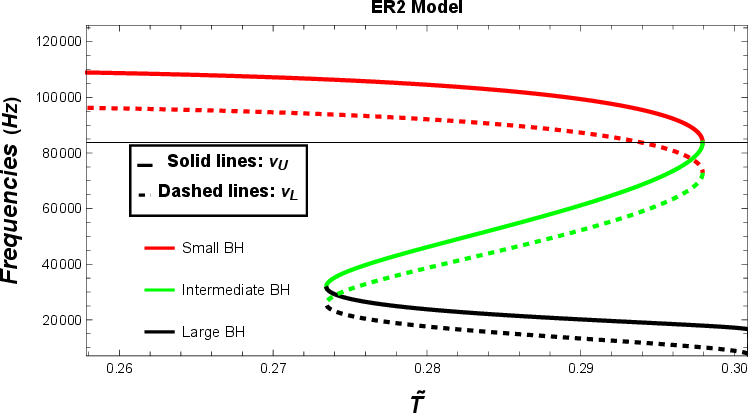}\hspace{1cm}
\includegraphics[width=7cm]{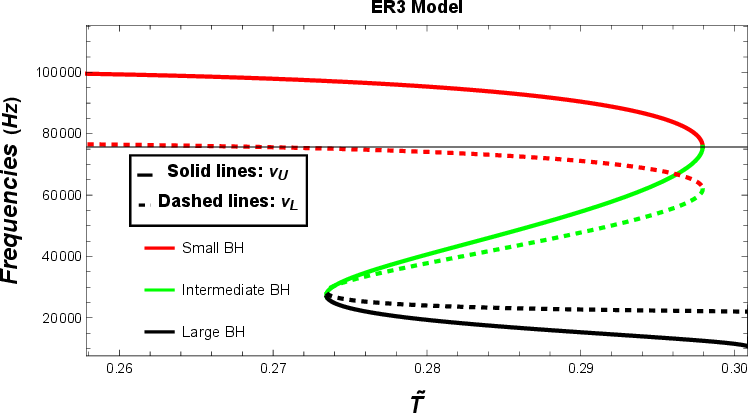}
\includegraphics[width=7cm]{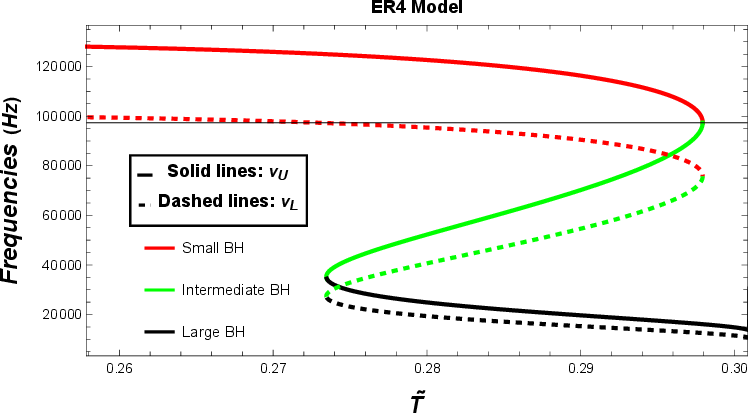}\hspace{1cm}
\includegraphics[width=7cm]{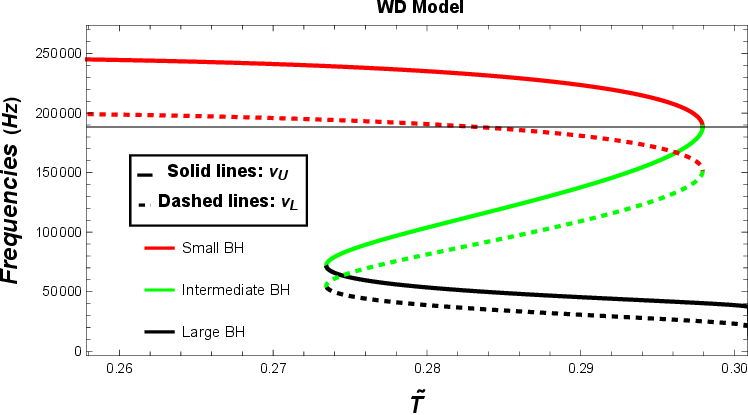} 
\includegraphics[width=9cm]{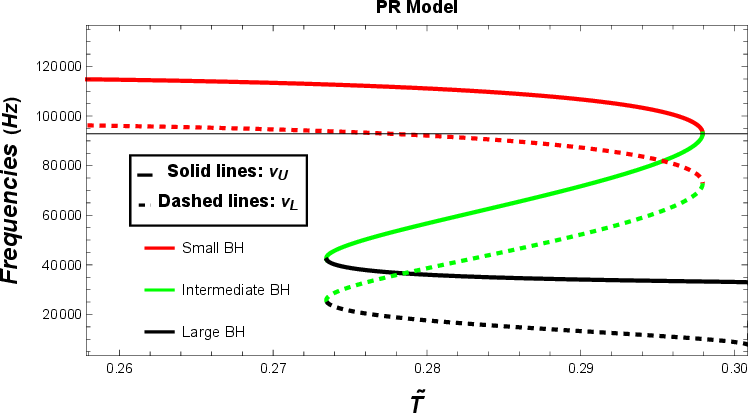}\hspace{1cm}

\caption{Upper $\left(\frac{2l}{r_s}\right)\nu_U$ and lower $\left(\frac{2l}{r_s}\right)\nu_L$ QPO frequencies as functions of the Hawking temperature for the ER2, ER3, ER4, WD, and PR models with fixed parameters $\tilde{Q}_{m}=0.03$, $\tilde{\xi}=0.15$ and $\tilde{L}=20$. In each model, the multivalued frequency--temperature relation reproduces the thermodynamic phase structure of the regular magnetic black hole, demonstrating that the thermodynamic imprints in the QPO spectrum are robust across different QPO prescriptions.} \label{fig:all models}
\end{figure}

To examine whether the thermodynamic imprints discussed in the main text persist across different QPO prescriptions, we consider several well established QPO models~\cite{Hazarika:2025zgv, Shahzadi:2021upd}. These include the epicyclic resonance (ER) models: ER2 ($\nu_U = 2\nu_{\theta} - \nu_r,\ \nu_L = \nu_r$), ER3 ($\nu_U = \nu_{\theta} + \nu_r,\ \nu_L = \nu_{\theta}$), and ER4 ($\nu_U = \nu_{\theta} + \nu_r,\ \nu_L = \nu_{\theta} - \nu_r$); the warped-disk (WD) model, defined by ($\nu_U = 2\nu_{\phi} - \nu_r,\ \nu_L = 2(\nu_{\phi} - \nu_r)$); and the parametric resonance (PR) model, given by ($\nu_U = \nu_{\theta},\ \nu_L = \nu_r$).
In these models, the observed upper and lower QPO frequencies are constructed from different combinations of the orbital, radial epicyclic, and vertical epicyclic frequencies of test particle motion in the black hole spacetime. Although the physical mechanisms underlying these models differ, they consistently exhibit the same qualitative thermodynamic signatures. The corresponding frequency--temperature profiles are shown in Fig.~\ref{fig:all models}, in which the behavior is similar to that in Fig.~\ref{fig:T-U}.

\bibliography{ref}
\bibliographystyle{utphys}

\end{document}